\documentclass[useAMS,usenatbib]{mn2e}
\usepackage{amssymb,amsmath,epsfig,times,float}

\title[Abell 2029 at the virial radius]{X-ray observations
  of the galaxy cluster Abell 2029 to the virial radius}
\author[S. A. Walker et al.]{S. A. Walker,$^1$\thanks{Email:
    swalker@ast.cam.ac.uk} A. C. Fabian,$^1$ J. S. Sanders$^1$, M. R.
George$^2$ and Y. Tawara$^3$\\
  $^1$Institute of Astronomy, Madingley Road, Cambridge CB3 0HA \\
  $^2$Department of Astronomy, University of California, Berkeley, CA 94720,
USA\\
  $^3$Department of Physics, Nagoya University, Nagoya, 338-8570, Japan
  \\
   \\
   \\
}
\date{}

\begin{document}

\maketitle

\begin{abstract}
We present \emph{Suzaku} observations of the galaxy cluster Abell 2029, which
exploit \emph{Suzaku}'s low particle background to probe the ICM at radii
beyond that which is possible with previous observations, and with better azimuthal coverage. We find significant
anisotropies in the temperature and entropy profiles, with a region of lower
temperature and entropy occurring to the south east, possibly the result of accretion activity in this direction. Away from this cold feature, the
thermodynamic properties are consistent with an entropy profile which rises,
but less steeply than the predictions of purely gravitational hierarchical
structure formation. Excess emission in the northern direction can be explained
due to the overlap of the emission from the outskirts of Abell 2029 and nearby
Abell 2033 (which is at slightly higher redshift).
These observations suggest that the
assumptions of spherical symmetry and hydrostatic equilibrium break down in the
outskirts of galaxy clusters, which poses challenges for modelling cluster
masses at large radii and presents opportunities for studying the formation and
accretion history of clusters.
\end{abstract}


\section{Introduction}
 \emph{Suzaku}'s ability to probe the intracluster medium (ICM) to the virial
radius has been shown for a number of relaxed clusters so far
(\citealt{George2009}; \citealt{Reiprich2009, Bautz2009,
Hoshino2010, Kawaharada2010, Simionescu2011}). The study of the outskirts of galaxy clusters has been prevented until recently due to its low surface brightness, and due to the high levels of background radiation to which X-ray telescopes are exposed. The virial radius broadly
represents the boundary between the central cluster which is in hydrostatic
equilibrium, and the outskirts where matter is still accreting onto the cluster
as it continues to form. These measurements are therefore crucial in
understanding the formation history of galaxy clusters, and in constraining
models which describe the formation process. They also allow mass estimates of
galaxy clusters to be improved by allowing the dynamic state of the ICM to be
understood to greater radii, whereas previous mass determinations have
extrapolated measurements of the centres of clusters. This increases the
accuracy with which galaxy clusters can be used as probes of cosmological
parameters (such as in \citealt{Allen2008}, and \citealt{Vikhlinin2009}). Here we add Abell 2029 to the growing collection of clusters which
have been investigated to the virial radius.

Abell 2029 (z=0.0767) is a relaxed galaxy cluster which has been studied
extensively in X-rays with \emph{ASCA} and \emph{ROSAT} \citep[][ hereafter
S98]{Sarazin1998}, BeppoSAX \citep{Molendi1999}, \emph{Chandra}
\citep{Lewis2002, Clarke2004, Vikhlinin2005, Vikhlinin2006}, and
\emph{XMM-Newton} \citep{Snowden2008, Bourdin2008}. It has a large cD galaxy
\citep{Uson1991} whose major axis is aligned in the NE to SW direction, in
approximately the same direction as that joining it to nearby Abell 2033.

\begin{figure}
  \begin{center}
    \leavevmode
      \epsfig{figure=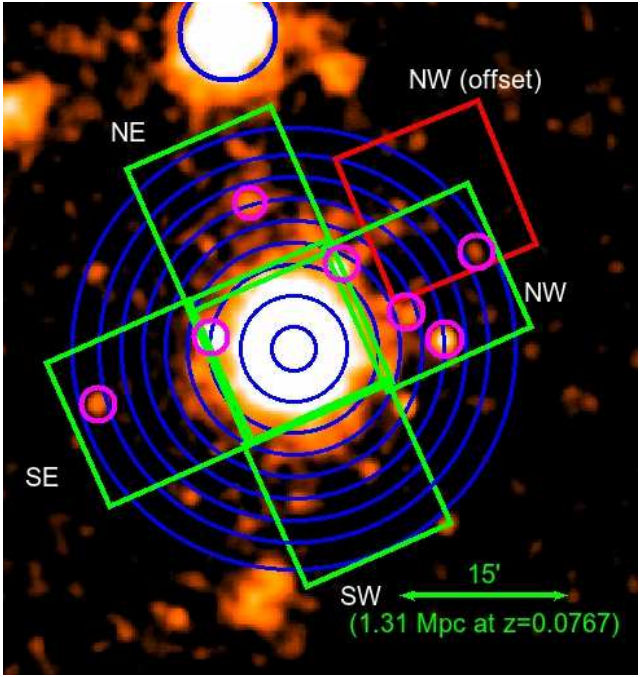,
        width=\linewidth}
      \caption{\emph{Suzaku} pointings of A2029 overlaid on a \emph{ROSAT} PSPC
image of Abell 2029. The five pointings are in green, the 6th pointing from the
public archive is in red. Abell 2033 is highlighted by the blue circle to the
North. The blue annuli shown are used in extracting spectra and have ring radii
at
2.5$'$, 6.0$'$, 9.5$'$, 12.0$'$, 14.5$'$, 17.0$'$, 19.5$'$, 22.0$'$ and
25.0$'$. Pointing identifiers are the same as those in Table \ref{obsdetails}.
Extracted point sources are shown circled in pink.  }
      \label{PSPC_pointing}
  \end{center}
\end{figure}

\begin{figure}
  \begin{center}
    \leavevmode
       \epsfig{figure=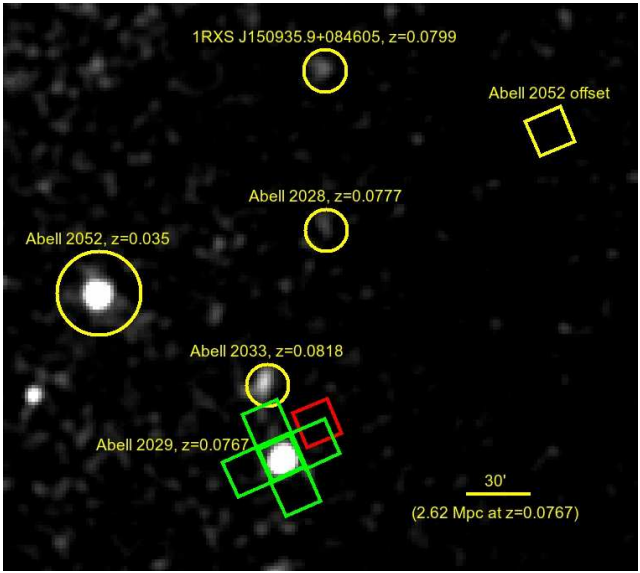, width=\columnwidth }
      \caption{Locations of galaxy clusters in the galaxy supercluster number
154 in the catalogue of \citet{Einasto2001}, shown on RASS hard band data. The
pointings for Abell 2029 are shown at the bottom of the image. Abell 2066 is
omitted as it lies ~6 degrees to the South East of Abell 2029. Abell 2052 is at
a different redshift and so not associated with the other clusters.}
      \label{RASS_supercluster}
  \end{center}
\end{figure}

Along with Abell 2033, Abell 2028 and Abell 2066, Abell 2029 forms a small
supercluster (catalogue number 154, \citealt{Einasto2001}), the closest cluster
of which is Abell 2033 at 35 arcmin to the North, as shown in Fig.
\ref{PSPC_pointing}. Abell 2029 provides an excellent opportunity to study how
the outskirts of a galaxy cluster are affected by its surroundings.
\citet{Gastaldello2010} has suggested that a further cluster, 1RXS
J150935.9+084605, is associated with the supercluster, and their locations are
shown on Fig. \ref{RASS_supercluster}.

\section{Observations and Data Reduction}
Five pointings in a cross formation were obtained during observations with
\emph{Suzaku} between 2010 January 28 and 30. A further pointing from the
public archive in the north west
direction was taken on 2008 January 8 and was also used. An overlay of these
pointings is shown in Fig. \ref{PSPC_pointing}, along with the annuli used in
the data analysis. The annular regions used lie between rings at radii 2.5$'$,
6.0$'$, 9.5$'$, 12.0$'$, 14.5$'$, 17.0$'$, 19.5$'$, 22.0$'$ and 25.0$'$.
The low exposure time of
the central pointing is compensated by the high surface brightness.

\begin{table*}
  \begin{center}
  \caption{Observational parameters of the pointings}
  \label{obsdetails}

%    \leavevmode
    \begin{tabular}{llllll} \hline \hline
    Name & Obs. ID & Position & Total Exposure (ks) & RA & Dec (J2000) \\ 
    \hline
    1 & 804024010 & Centre & 14.2  & 227.7409 & 5.7498 \\
    2 & 804024020 & NE & 43.4 &  227.8527 & 6.0108\\
    3 & 804024030 & SW & 33.4 &  227.6311 & 5.4874 \\
    4 & 804024040 & SE & 40.4 & 228.0053 &  5.6496\\
    5 & 804024050 & NW & 36.0 &  227.4808 & 5.8608 \\
    6 & 802060010 & NW (offset) & 289.2 & 227.4644 & 6.0238 \\
    \hline
    \end{tabular}
  \end{center}
\end{table*}

We use HEASoft version 6.11.1, and the calibration database (CALDB) released on
2011 November 10. The raw unfiltered XIS data were cleaned using the
\textsc{aepipeline} tool which performs
the standard screening described in Tables 6.1 and 6.2 of the \emph{Suzaku} ABC
guide\footnote[1]{http://heasarc.gsfc.nasa.gov/docs/suzaku/analysis/abc/}, and
in addition to the default screening criteria we use COR$>$6 GeV to restrict
the geomagnetic cut off rigidity of the particle background, as is typically
done \citep{Bautz2009, Kawaharada2010}. Calibration regions and bad columns
were then removed. The defective region in the XIS0 data (defective due to a
micrometeorite hit) was excluded. The observations were checked to ensure they weren't contaminated by solar wind charge exchange (SWCX) as described in section \ref{SWCX}, and the light curves of the observations were checked to ensure that no flaring had occured. 
Following the application of the all of this filtering, the good exposure times
for
each pointing are shown in Table \ref{obsdetails}.

Ancillary response functions (ARFs) and redistribution matrix files (RMFs) are
needed to model the response. These were created using the
\textsc{xissimarfgen} \citep{Ishisaki2007} and \textsc{xisrmfgen} ftools,
respectively.
The emission we detect can be divided into two broad components; the cluster
emission which varies with position on the sky, and the background which is
assumed constant on
the sky. Since the spatial variation is different for these two components,
their responses will be different. We model the response for the background by
assuming
a uniform background circular region of radius 20$'$ [beyond the size of the
XIS field of view (FOV)]. For the cluster emission we use as input to
\textsc{xissimarfgen} a model
of the surface brightness profile of the cluster. The model we use is the beta
model from S98 (their equation 5) obtained using \emph{ROSAT} data in the
0.5-2.0 keV band out to 17$'$, though we note that this fit is not
appropriate in the central region (r$<$164$"$) due to possible cooling flow
regions, and that here the surface brightness is higher than the model
predicts. This model has also been found to be consistent with \emph{ASCA} data
in both the 0.7-1.5 keV and 3-10 keV bands \citep{Miyoshi2005}, and we show
later in section \ref{Filament} that it is consistent with our \emph{Suzaku} data. For
the central region (r$<$164$"$) we use the beta model found in
\citet{Lewis2002} using \emph{Chandra} data for the surface brightness model,
and match the two beta models at r=164".

It has been found previously \citep{George2009} that the results are relatively unaffected
by changing the surface brightness distribution input to \textsc{xissimarfgen},
so we do not expect possible deviations from the $\beta$ model in the outskirts to affect the results significantly. Indeed, when the
analysis was performed using only the S98 $\beta$ model as the surface
brightness input to \textsc{xissimarfgen}, (which underestimates the surface
brightness of the core) the results of the spectral fits were essentially
unchanged.

We use only data from the front illuminated (FI) detectors XIS0 and XIS3 owing
to the higher background level of the back illuminated (BI) detector XIS1
\citep{George2009}. We use both the
3$\times$3 editing mode and the 5$\times$5 editing mode. The spectral fits are
performed by fitting the data for each detector and editing mode
simultaneously.
We use a standard $\Lambda$CDM cosmology with $H_{0}=70$  km s$^{-1}$
Mpc$^{-1}$, $\Omega_{M}=0.3$, $\Omega_{\Lambda}$=0.7. All errors unless
otherwise stated are at the 1 $\sigma$ level.

All spectral fits were performed in \textsc{xspec} 12.7.0u using the extended C-statistic.

\section{Background Modelling and Subtraction}
\label{backgroundmodelling}

Accurate measurements of the low surface brightness outskirts of galaxy
clusters requires an accurate understanding of the X-ray background and its
expected level of variation. The X-ray background can be
separated into two main distinct components. One is the cosmic X-ray background
(CXB) which consists of unresolved point sources of extragalactic origin, mostly
believed to be AGN, and which dominates the background in the
2-10keV band. The second is soft foreground emission from within our galaxy
which is the dominant contributor to the X-ray background at energies below
2 keV; this is the result of hot gas within our galaxy
and so the nature of this background component is strongly dependent on the
direction of observation. We treat each of these
background components separately as follows.

 \subsection{The CXB}
\label{CXB_level}

The CXB has been studied extensively by most major X-ray satellites, however
measurements of its average flux in the 2-10 keV band have varied between
different observatories, as shown in Fig. 10 of \citet{DeLuca2004}.
We take the total CXB flux in the 2-10 keV band to be the value from
\citet{Moretti2009} using \emph{Swift} data, 2.18 $\pm$0.13 $\times$ 10$^{-11}$ erg
cm$^{-2}$ s$^{-1}$ deg$^{-2}$, which is highly consistent with the value
obtained using \emph{XMM-Newton}
(2.24 $\pm$ 0.16 $\times$ 10$^{-11}$ erg cm$^{-2}$ s$^{-1}$ deg$^{-1}$ in
\citealt{DeLuca2004}). These values are higher than those obtained using ASCA in
\citet{Kushino2002} which are used to model the CXB in \citet{Hoshino2010} and
\citet{Akamatsu2011} in their studies of \emph{Suzaku} data for the outskirts of Abell
1413 and Abell 2142 respectively, which is most likely due to the significant
stray light contamination to which the ASCA observations were subjected.

Our \emph{Suzaku} observations allow us to exclude point sources down to a threshold
flux of S$_{excl}$ = 10$^{-13}$ erg cm$^{-2}$ s$^{-1}$ in the 2-10 keV band
(and the point sources excluded are shown in Fig. \ref{PSPC_pointing}). To
calculate the unresolved CXB level following the removal of these point
sources, we use the source flux distribution found in \citet{Moretti2003}
(obtained by fitting to a catalogue of point sources extending down to a flux
of 2 $\times$ 10$^{-16}$ erg cm$^{-2}$ s$^{-1}$) to calculate
the flux contribution of the resolved sources. We do this instead of
integrating the source flux distribution from 0 to the threshold flux because
the source flux distribution is uncertain
below the minimum flux studied in \citet{Moretti2003}, and indeed when
integrated over all energies leads to an underestimate of the total CXB level
(see Fig. 5 of \citealt{Moretti2003}).

The unresolved CXB level (in ergs cm$^{-2}$ s$^{-1}$ deg$^{-2}$) is therefore
given by:

\begin{eqnarray}
F_{\rm CXB} = 2.18 \pm 0.13 \times 10^{-11} - \int_{S_{\rm excl}}^{S_{\rm max}}
\Big(\frac{dN}{dS} \Big)
\times S ~ dS
\end{eqnarray}

which, using the two powerlaw best fit for N(S) given in \citet{Moretti2003}
gives an unresolved 2-10 keV flux of  1.87$\pm$0.13 $\times$ 10$^{-11}$ erg
cm$^{-2}$ s$^{-1}$ deg$^{-2}$.

The expected fluctuations in the CXB for a given solid angle, $\Omega$, due to variations
in the number of unresolved point sources below the threshold flux in the area
of observation can be calculated using \citep{Bautz2009}:

\begin{eqnarray}
\sigma^2_{\rm CXB} = (1/\Omega) \int_{0}^{S_{\rm excl}} \Big(\frac{dN}{dS}
\Big)
\times S^2 ~ dS
\end{eqnarray}

where we see that the variance decreases with increasing solid angle observed
($\Omega$) and with decreasing threshold flux (S$_{\rm excl}$).

The expected fluctuations in the CXB for each extraction region examined are
shown in table \ref{CXB_fluctuations}. These provide the ranges which must be
investigated to understand the effect of the
uncertainty of the CXB level on our measurements in the outskirts. These ranges
are consistent with those obtained by scaling the fluctuations observed with
Ginga using the method of \citet{Hoshino2010}, and also with the observed
variance described in \citet{Simionescu2011conf} using the \emph{Suzaku} background
pointings for the Perseus cluster.

\begin{table*}
  \begin{center}
  \caption{CXB fluctuations in each annulus for each pointing in the 2-10 keV band (10$^{-12}$ erg s$^{-1}$
cm$^{-2}$ deg$^{-2}$)}
  \label{CXB_fluctuations}

%    \leavevmode
    \begin{tabular}{llllllllll} \hline \hline

       Position         &     0.0$'$-2.5$'$   &  2.5$'$-6.0$'$  & 6.0$'$-9.5$'$   &  9.5$'$-12.0$'$
&12.0$'$-14.5$'$   &  14.5$'$-17.0$'$  &   17.0$'$-19.5$'$ &  19.5$'$-22.0$'$ & 22.0$'$-25.0$'$    \\ 
\hline
    Centre & 6.8   &   3.1  &    2.7   &       -    &      -     &
-   &       -    &      -   &       - \\
    NE &       -  &    -     &     -  &    6.1   &   5.0  &    4.6
&    4.5  &   4.5  &    4.6\\
    SW &   -& -    &      -    &  5.6 &    5.0   &   4.6 &
4.5  &    4.6  &    4.9\\
    SE &   - & -   &       -  &    5.5 &     4.6 &     5.0  &
4.7 &    5.2 &    6.4\\
    NW &   - & -    &      -   &   6.6  &    6.5   &   4.9  &
5.1 &    5.3 &     5.0 \\
    NW (offset) &   -& -    &      -     &     -       &   -   &   8.2 &
5.6 &    5.7  &    5.1\\
\hline
    \end{tabular}
  \end{center}
\end{table*}

\subsection{The soft galactic foreground}
\label{GAL_level}

Abell 2029 is located well out of the galactic plane, meaning that the
absorbing hydrogen column density is low [the mean Leiden Argentine Bonn (LAB)
survey \citep{LAB2005} value is 3.26$\times$ 10$^{20}$ cm$^{-2}$] and that
possible variations of the column
density across the cluster have a negligible affect on the soft band emission.

Abell 2029 is however located near the North Polar Spur (NPS), which is a
region of complex soft band emission attributed to part of a hot interstellar
bubble formed from supernova explosions and the outflows of young, hot stars.
Previous studies of Abell 2029 and its surrounding
region have required extra background components to model the NPS emission, in
addition to the background level associated with normal galactic emission.
Studies of Abell 2029 with \emph{Chandra} \citep{Vikhlinin2005} and with
\emph{XMM-Newton} \citep{Bourdin2008} have used a soft foreground model consisting of
two unabsorbed \textsc{mekal} components, one at 0.46 $\pm$ 0.07 keV and the other at 0.22 $\pm$ 0.05 keV.

We use the region outside 25$'$ from the north-western offset \emph{Suzaku} pointing to fit
for the soft thermal background components, and find that that the best fit
(shown in Appendix A) is achieved using a model consisting of two
unabsorbed \textsc{apec} components with temperatures at 0.53$^{+0.08}_{-0.08}$ keV and
0.20$^{+0.05}_{-0.05}$ keV, each of which has its metallicity fixed to 1 Z$_{\odot}$. This is consistent with the foreground model used in \citet{Vikhlinin2005} described earlier, which themselves were found to be consistent with previous \emph{Chandra} measurements of the NPS region. 
To investigate the spatial variation of these background components, we use
the ROSAT PSPC pointing of Abell 2029 (observation identifier: rp800249) and extract
four background regions (in an annulus between 30$'$ to 50$'$) to the northwest, sorthwest,
northeast and southeast, and fit the spectra in the 0.4-2.0keV bands after
subtracting the particle background as modelled by the ftool \textsc{pcparpha}. ROSAT ARFs were created using the ftool \textsc{pcarf} and standard filtering procedures were used as described in the document 'ROSAT data analysis using xselect and ftools'  \footnote[2]{heasarc.gsfc.nasa.gov/docs/rosat/ros\_xselect\_guide/xselect\_ftools.html}. We find that the
spectra are completely consistent with the two \textsc{apec} component model found
using the \emph{Suzaku} region to the north-west, and table \ref{GALvariations} shows
the spatial variations in the normalisations of
the two background components, which we use as a indicator of the extent of
expected spatial variations of these soft background components. The fits to
the ROSAT data are also shown in Appendix A. These variations will be used
later,
along with the variations
expected in the CXB from table \ref{CXB_fluctuations}, to determine the affect
on our fits of our uncertainty in the background model.

When fitting for the soft foreground the CXB is modelled as an absorbed powerlaw of index 1.4 and its normalisation is allowed 
to be a free parameter. In all cases the CXB normalisation is consistent with the expected level given the level of point source removal. 

\begin{table*}
  \begin{center}
  \caption{Soft foreground fluctuations from ROSAT PSPC and \emph{Suzaku} data. The units of the \textsc{apec} normalisations are $10^{-14}(4\pi)^{-1}D_{A}^{-2}(1 + z)^{-2} \int n_{e} n_{H} dV$, where   $D_{A}$ is the angular size distance (cm), and $n_{e}$ and $n_{H}$ are the electron and hydrogen densities (cm$^{-3}$) respectively, and these values are scaled for a circular area of sky of 20$'$ radius (1257 arcmin$^{2}$).}
  \label{GALvariations}

    \leavevmode
    \begin{tabular}{llllllllll} \hline \hline

       Position   &  0.53 keV \textsc{apec} norm & 0.20 keV \textsc{apec} norm      \\ \hline
    ROSAT NE &  $5.5^{+0.6}_{-0.6}$ $\times$ 10$^{-4}$ & $3.0^{+0.1}_{-0.1}$
$\times$ 10$^{-3}$ \\
    ROSAT SE  & $6.1^{+0.7}_{-0.7}$ $\times$ 10$^{-4}$ & $3.6^{+0.2}_{-0.2}$
$\times$ 10$^{-3}$ \\
    ROSAT NW &  $6.1^{+0.5}_{-0.5}$ $\times$ 10$^{-4}$ & $2.9^{+0.1}_{-0.1}$
$\times$ 10$^{-3}$ \\
    ROSAT SW  & $6.9^{+0.9}_{-0.9}$ $\times$ 10$^{-4}$ & $4.3^{+0.3}_{-0.3}$
$\times$ 10$^{-3}$ \\
    SUZAKU NW offset (25$'$-29$'$)  & 6.8$^{+0.3}_{-0.3}$ $\times$ 10$^{-4}$ &
3.5$^{+0.3}_{-0.3}$ $\times$ 10$^{-3}$\\
    SUZAKU NW offset (29$'$-34$'$)  & 6.7$^{+0.5}_{-0.5}$ $\times$ 10$^{-4}$ &
3.4$^{+0.3}_{-0.3}$ $\times$ 10$^{-3}$\\ \hline
    \end{tabular}
  \end{center}
\end{table*}

\subsection{Non X-ray Background}
\label{NXB}
We model the non X-ray background (NXB) resulting from interactions with
charged particles using the \textsc{xisnxbgen} ftool, using a database of
measurements taken in an interval spanning 150 days before and 150 days after
the observations, and subtract this from the spectra.

\subsection{Stray light modelling.}
\label{stray_modelling}

Due to broad PSF each annulus contains contributions from neighbouring annuli.
In addition, the precollimator is not completely successful in preventing light
following undesired paths (such as back reflection and secondary reflection, as described in 
\citealt{Serlemitsos2007}), and so stray light enters each pointing from the
bright core.

To understand the magnitude of this effect we use the image of each annulus
(taken to be represented by the beta model of S98 which we show later fits our
\emph{Suzaku} data well) as input to the ray tracing simulator \textsc{xissim} and simulate a 2$\times$10$^{6}$ photon exposure for
each annulus, and then measure the fraction of photons from each annulus which
are spread into the other annuli. The resulting distribution of photons is
shown in table \ref{PSF_spread_table}. We see that the majority of the flux detected in each annulus originated from that annulus, and that in the outskirts the
majority of the contamination originates from the nearest innermost annulus
which has similar properties to the annulus in question. 

We model emission from the central 6.0$'$ (from which the majority of the cluster
flux originates) which spreads into the arms of the cross using \textsc{xissim}, using
the spectra and image of the core as inputs. We model this emission and include
it as a background component, and investigate the affect of accounting for this
stray light (i.e. the systematic error caused by the stray light) on our fits
in section \ref{errors}.

\begin{table*}
  \begin{center}
  \caption{Percentage contribution of flux in the rows' annulus from the
columns' annulus due to PSF spreading and stray light.}
  \label{PSF_spread_table}

    \leavevmode
    \begin{tabular}{lllllllll} \hline \hline

 &  0.0$'$-2.5$'$   &  2.5$'$-6.0$'$  & 6.0$'$-9.5$'$   &  9.5$'$-12.0$'$
&12.0$'$-14.5$'$   &  14.5$'$-17.0$'$  &   17.0$'$-19.5$'$ &  19.5$'$-22.0$'$
\\ \hline
0.0$'$-2.5$'$      &  95.9  &    3.9  &  0.05 &  0.002 &
0.0002 & 0.0004 & 0.0004 & 0.0001\\
2.5$'$-6.0$'$      &  36.2  &    60.2  &    3.5 &   0.04 &
0.009 &  0.004 &  0.003 &  0.002\\
6.0$'$-9.5$'$      &  14.5  &    19.8  &    64.4 &     1.1  &
0.07  &  0.02 &   0.02 &  0.01\\
9.5$'$-12.0$'$      &  6.7 &    3.5  &    15.9 &    63.7  &
9.1  &  0.7  &   0.2  &  0.05\\
12.0$'$-14.5$'$      &  7.6  &    3.5  &   4.0  &   13.8    &
62.5   &  7.3   &  0.8   &  0.2\\
14.5$'$-17.0$'$      &  4.3  &    1.6  &    2.3 &      1.9 &
15.4  &    62.8  &    11.1 &   0.6\\
17.0$'$-19.5$'$      &  1.3  &    1.0  &   1.3  &    1.1   &
2.0   &   19.8   &   65.8  &   7.4\\
 19.5$'$-22.0$'$     &   5.2 &    1.2  &    1.0  &   0.7  &
1.1   &   3.0  &    19.8  &    65.1\\ \hline

    \end{tabular}
  \end{center}
\end{table*}

\begin{figure*}
  \begin{center}

    \leavevmode
        \hbox{
       \epsfig{figure=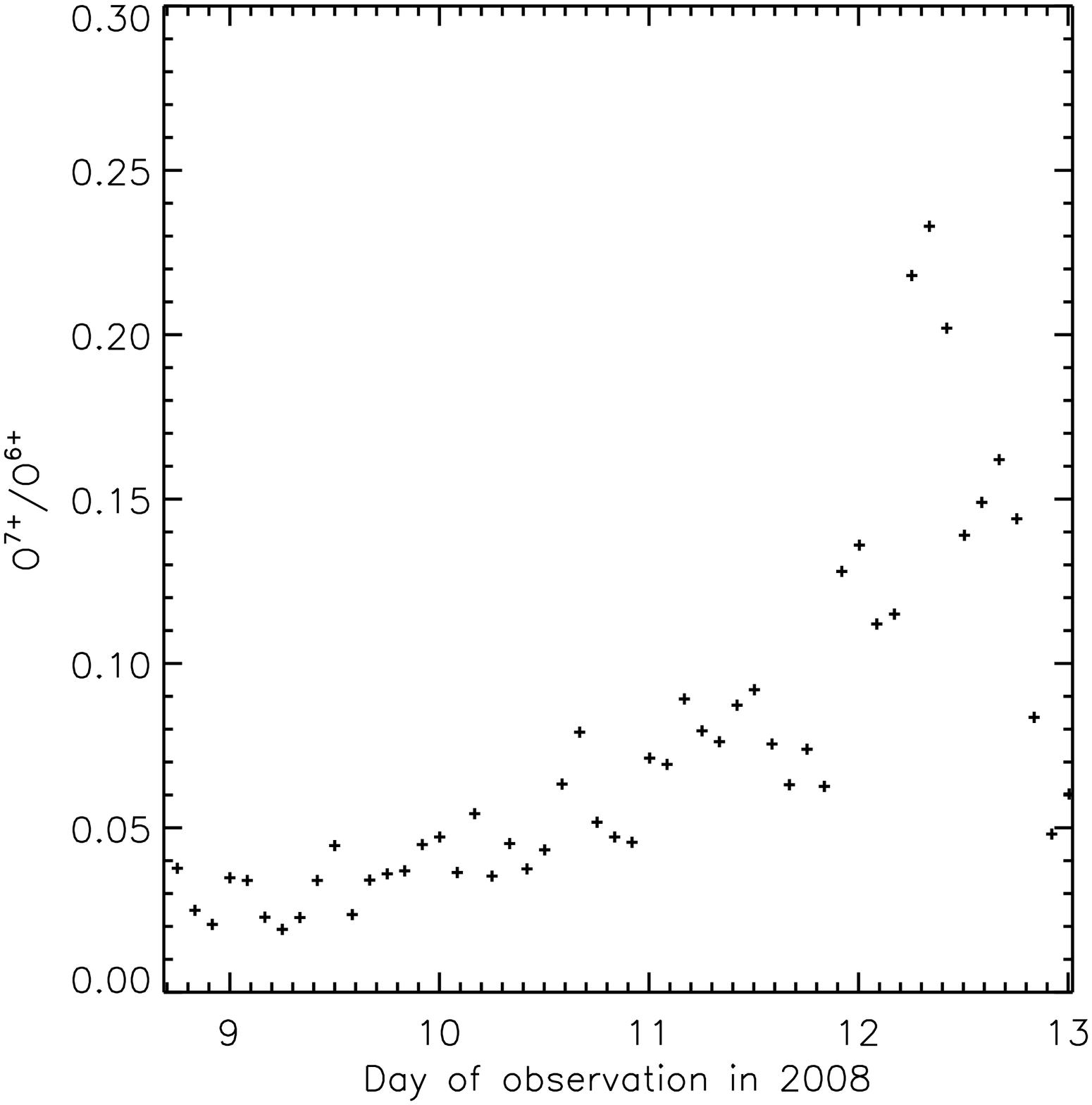, width=\columnwidth }
       \epsfig{figure=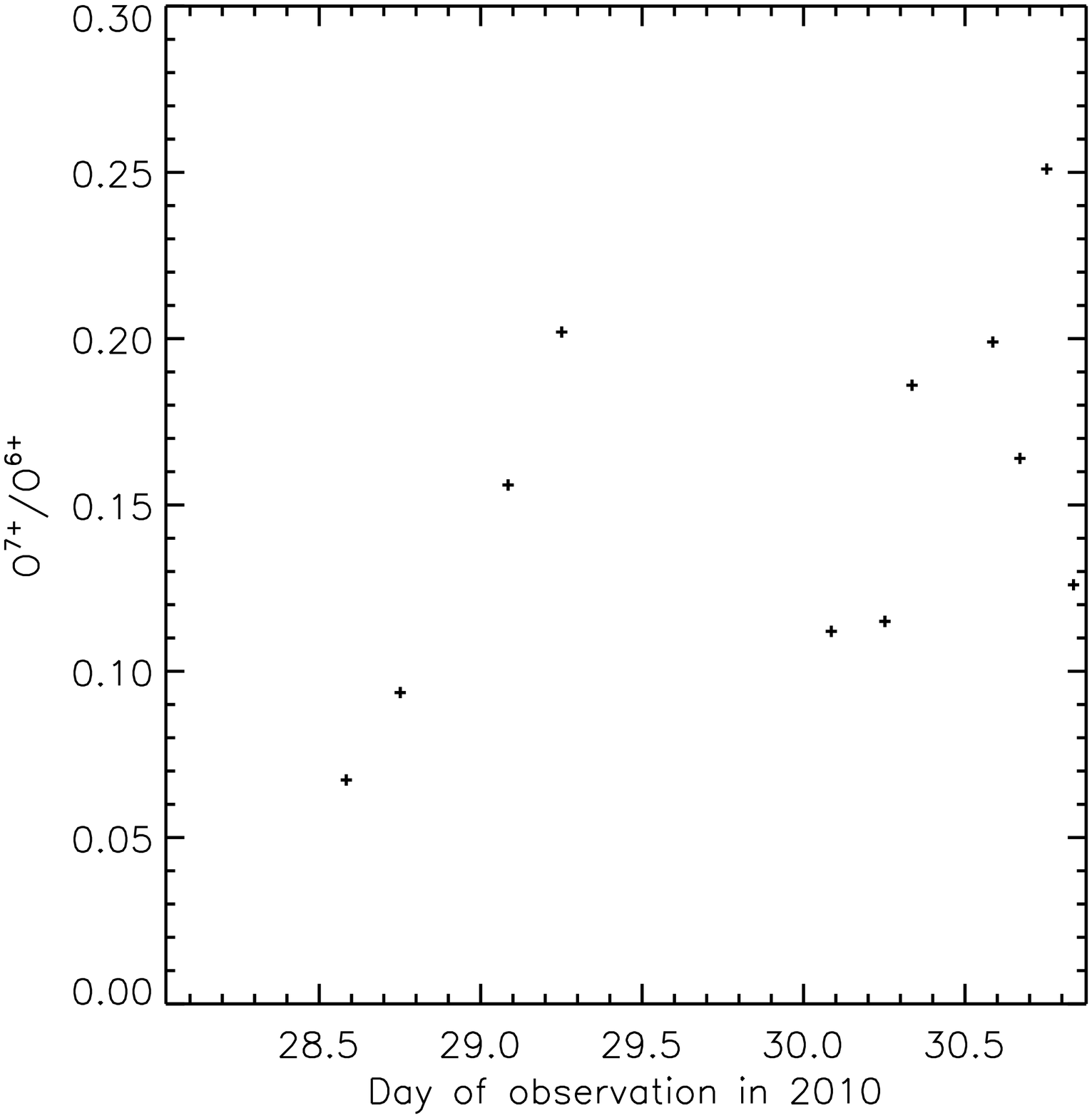, width=\columnwidth }
       }
      \caption{O$^{7+}$/O$^{6+}$ ratio as measured by the ACE spacecraft. SWCX
is assumed to be negligible when O$^{7+}$/O$^{6+}$ is below $\sim$ 0.2, and
regions above this level are excluded. Left is for the observation taken in
2008 (802060010), and right is for the time region of the observations in 2010
(804024010, 804024020, 804024030, 804024040, 804024050). The day of observation time axis label is the day of the year on which the observation took place. }
      \label{SWCXfigure}
  \end{center}
\end{figure*}

\subsection{Solar Wind Charge Exchange}
\label{SWCX}

To ensure our observations are not significantly contaminated due to solar wind
charge exchange (SWCX) emission we use data from the Solar Wind Ion Composition
Spectrometer (SWICS) on the Advanced Composition Explorer (ACE) spacecraft. We
follow \citet{Snowden2004}, which assumes that the SWCX is negligible when the
ratio of O$^{7+}$ to O$^{6+}$ is below $\sim$ 0.2 (this criterion is also used
in \citealt{Humphrey2011}), and we show in figure \ref{SWCXfigure} that only
very small time intervals need to be excluded because of this, and that
O$^{7+}$/O$^{6+}$ does not increase substantially above 0.2 during the times of
observation.

\section{Modelling cluster emission}
\label{Spectral_fits}

We fix the X-ray background model using the CXB level determined in section
\ref{CXB_level} following point source subtraction and model the CXB as an absorbed powerlaw of index 1.4,
added to the galactic components (two \textsc{apec} components) found in section
\ref{GAL_level} by fitting to the outer annuli of the north western offset
pointing (which are found to be consistent with ROSAT results for a background annulus
between 30$'$-50$'$ around the cluster). Later we will investigate the systematic
error of fixing the background by varying its parameters within the confidence
regions we determined in section \ref{backgroundmodelling}.

\subsection{Projected fits}
\label{Projected_fits}
First of all we do projected fits, modelling each annulus as an \textsc{apec} component
absorbed by a column density of 3.26$\times$10$^{20}$ cm$^{-2}$ (from the LAB survey) using the \textsc{phabs} absorption model.
We fit to 0.7-7.0 keV band and for the central 3 annuli we allow
the metallicity to be a free parameter. In the 0.0$'$-2.5$'$, 2.5$'$-6.0$'$ and 6.0$'$-9.5$'$ annuli the best fit projected metallicities are 0.45$^{+0.03}_{-0.03}$ Z$_{\odot}$, 0.3$^{+0.08}_{-0.08}$ Z$_{\odot}$ and 0.2$^{+0.1}_{-0.1}$ Z$_{\odot}$, which are reasonably consistent with the metallicities found in \citet{Vikhlinin2005}. In the annuli outside 9.5$'$ the
metallicity is fixed to 0.1 Z$_{\odot}$ as it is not constrained in fits (this is the
value obtained in \citealt{Vikhlinin2005} in the outskirts). We find that varying the metallicity in the outskirts through a reasonable range (from 0 Z$_{\odot}$ to 0.3 Z$_{\odot}$) has a negligible affect on the fitting parameters. We fix the redshift
to 0.0767.
We tie the temperatures of the outer 3 annuli due
to the low number of counts in the outermost annuli, but let their normalisations
be free between them.

We initially examined each direction separately and searched for significant
azimuthal variations which would make averaging over certain directions
inappropriate. We find two main sources of asymmetry: firstly the temperature
is lower in the south eastern direction than in the other directions which are
consistent with one another. This is shown in Fig. \ref{deprojected_fits} (the
deprojected temperatures shown there are consistent within errors with the
projected temperatures, which has also been observed in \citealt{Bautz2009} and \citealt{Akamatsu2011}). This lower temperature to the south east coincides
with a depression in temperature seen in the temperature map produced using \emph{XMM-Newton}
data in \citealt{Bourdin2008} to the south east, suggesting a disturbance of
the ICM in this
direction extending from the core of the cluster all the way to the outskirts,
possibly due to the accretion of small galaxy groups in this direction.

\begin{figure}
  \begin{center}
    \leavevmode
        \hbox{
       \epsfig{figure=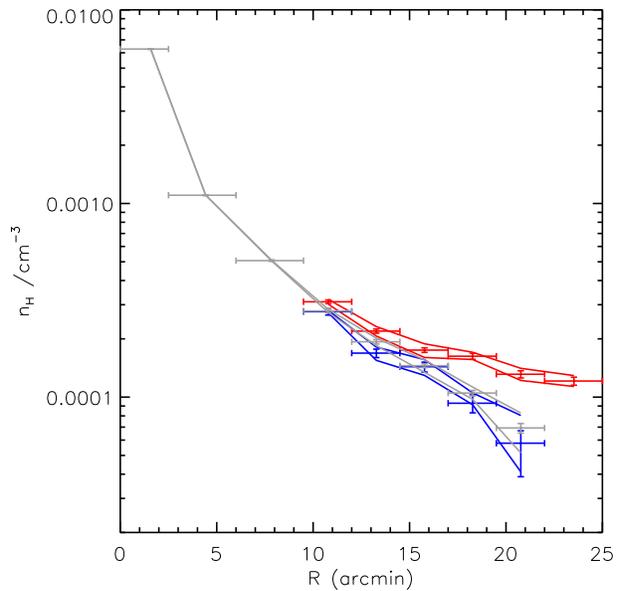,
width=\columnwidth }
       }
      \caption{Comparing projected densities. Grey is the azimuthal average excluding the north. Blue is for the low temperature south eastern direction. Red is for
      the northern direction, showing a significant excess. Solid lines show 1 $\sigma$ systematic errors calculated as described in section \ref{errors}.}
  \label{projected_densities}
  \end{center}
\end{figure}

Secondly the projected density is highest to the north and flattens off (see
Fig. \ref{projected_densities}) and this is the only direction in which we detect statistically significant emission beyond 22$'$. This indicates the presence of excess emission
above the azimuthal average which may
indicate a filament structure to the north connecting Abell 2029 and Abell
2033, or emission from Abell 2033 entering the field of view. We investigate this in depth in section \ref{Filament}.

\subsection{Deprojected fits}
\label{Deprojected_fits}
We perform two deprojection runs. The first averages over all of the directions
except for the northern
direction (where excess emission exists, rendering deprojection impossible in
this direction), and the south eastern direction (which has a lower
projected temperature and entropy than the other directions). In the second the
south eastern direction is
deprojected individually due to its lower temperature and entropy.

To achieve the deprojection we fit all nine annuli simultaneously, with each
annulus modelled as the
superposition of the emission from the shell associated with the annulus and
the
emission projected into that annulus from shells exterior to it under the
assumption of spherical symmetry (thus emulating the \textsc{projct} mixing model by adding multiple \textsc{apec} components as is done in \citealt{Humphrey2011}, which allows data from multiple \emph{Suzaku} instruments and pointings to be fitted simultaneously). The temperature is tied for the 3 annuli
between 17' and 25' and for the 3 annuli between 9.5' and 17' due to the
limited spectral quality of the data. The resulting deprojected temperatures
are consistent with the projected ones. The deprojected temperature, density
and entropy (S=kT/n$_{e}^{2/3}$) profiles are shown in Fig. \ref{deprojected_fits} for the azimuthal
average and the south eastern direction separately.

The lower temperature in the south eastern direction causes the entropy profile to
flatten indicating significant deviation from hydrostatic equilibrium.
The temperature profile to the south east maps on well to the temperature profile
found in \citet{Bourdin2008} for the low temperature region they find to the south
east (within 8 arcmins of the core), and is overplotted in blue in Fig.
\ref{deprojected_fits}. The lower entropy profile from \emph{XMM-Newton} in the
central 8 arcmins also maps on well to the flatter and lower entropy we find in
the outskirts to the south east, suggesting that the ICM has been disturbed from
equilibrium from the outskirts to near
the core, possibly due to the accretion of small galaxy groups along this
direction (as was suggested in \citealt{Bourdin2008}).

Due to \emph{Suzaku's} large PSF it is not possible to divide the inner two annuli into sectors corresponding 
to the arms of the cross, but when the 3rd annulus (between 6.0$'$-9.5$'$) is divided we find a statistically significant drop in the temperature below the 
azimuthal average, which is consistent with the \emph{XMM-Newton} temperature in the same region, as shown in the left column of Fig. \ref{deprojected_fits}. 

 For the azimuthally averaged fits the entropy profile rises with increasing
radius, but less steeply than the r$^{1.1}$ relation derived in
\citet{Voit2005} assuming purely gravitational hierarchical structure formation
(and neglecting non gravitational processes such as feedback, cooling and star
formation).

\begin{figure}
  \begin{center}

    \leavevmode
        \hbox{
       \epsfig{figure=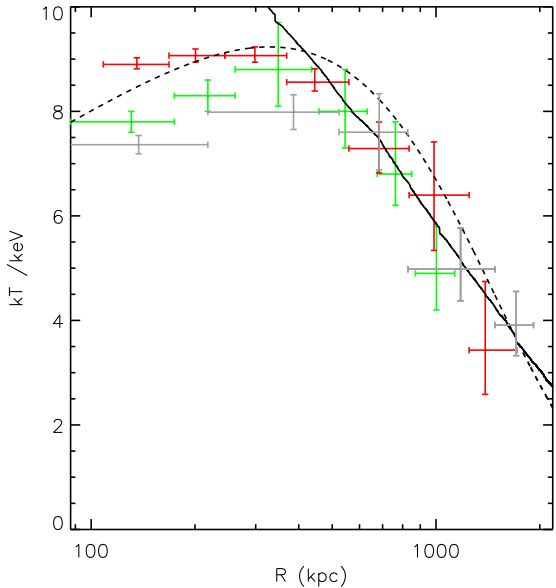, width=\columnwidth }
       }
      \caption{Comparing \emph{Suzaku} temperature profiles with projected
temperatures obtained with \emph{Swift} in \citet{Moretti2011} (green) and \emph{Chandra}
(red) in \citet{Vikhlinin2006}. Grey crosses are azimuthally averaged \emph{Suzaku}
temperatures, (exluding the north and the south east). The black
dashed line is the deprojected temperature from \emph{Chandra} in \citet{Vikhlinin2006}
and the solid black line shows the best fit profile from 6$'$ to 22$'$ to the deprojected
\emph{Suzaku} temperatures.}
      \label{temperature_compare}
  \end{center}
\end{figure}

The temperature profile for all directions neglecting the south east is consistent
with what has previously been found using \emph{Chandra}, \emph{Swift} and \emph{XMM-Newton}, as
shown in Fig. \ref{temperature_compare} and Fig. \ref{deprojected_fits}. The deprojected density profile in all directions is consistent with the
extrapolated best fit
model to \emph{Chandra} data in \citep{Vikhlinin2006} when systematic errors in the
background modelling are taken into account, as shown in Fig. \ref{deprojected_fits}. 

\subsection{Systematic errors}
\label{errors}
We need to understand the contribution of the uncertainty of the background
model to the error budget. The normalisations of the CXB and the galactic
components, (and the temperatures of the galactic components) are only known to an accuracy described in section
\ref{backgroundmodelling}, so we need to see the effect on our fits by
exploring the possible background parameter space. Typically (for example
\citealt{George2009} and \citealt{Akamatsu2011}), these systematic errors have been
quantified by varying only one parameter at a time and leaving the others
unchanged, which underestimates the actual uncertainty.

\begin{figure*}
  \begin{center}

    \leavevmode
        \hbox{
       \epsfig{figure=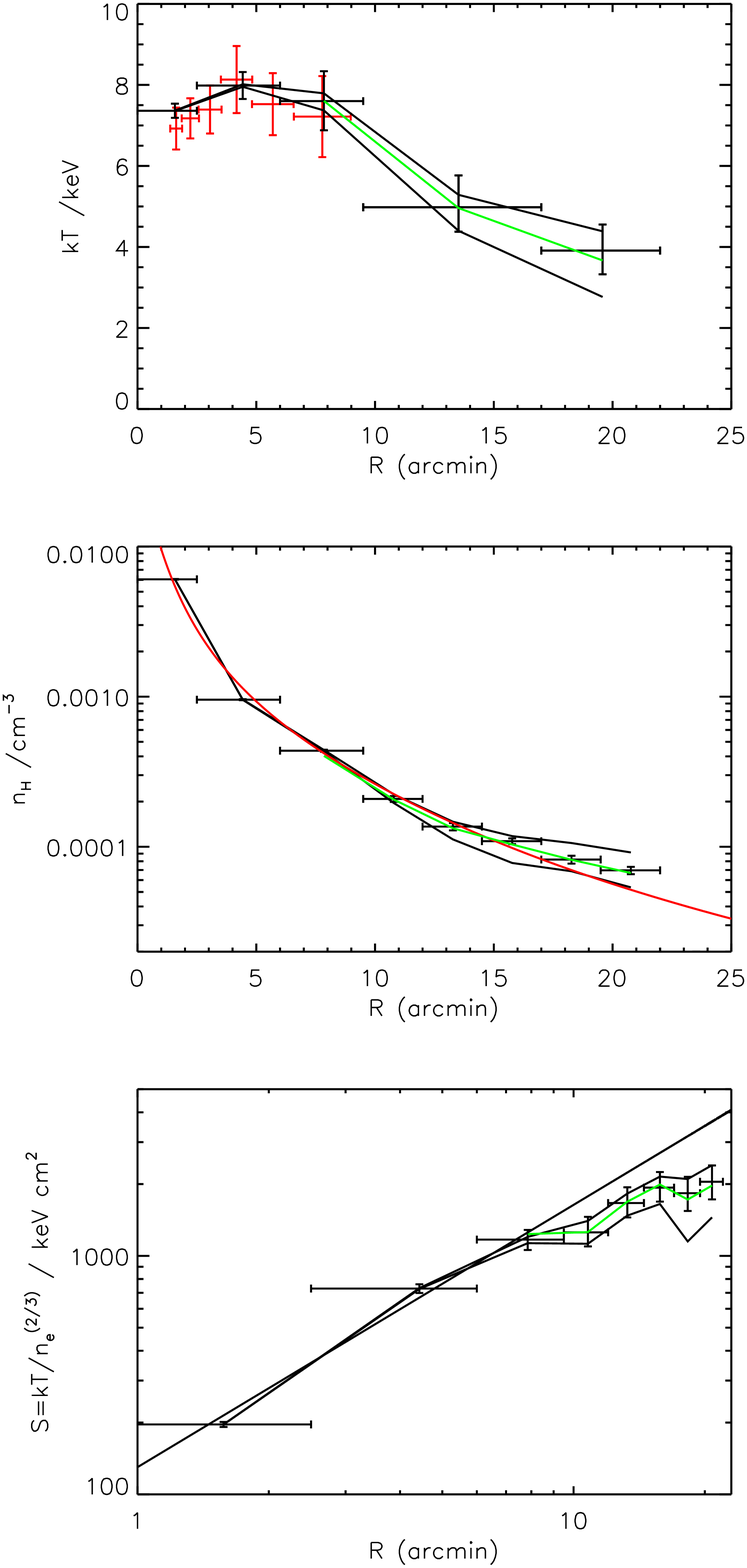,
width=\columnwidth }
       \epsfig{figure=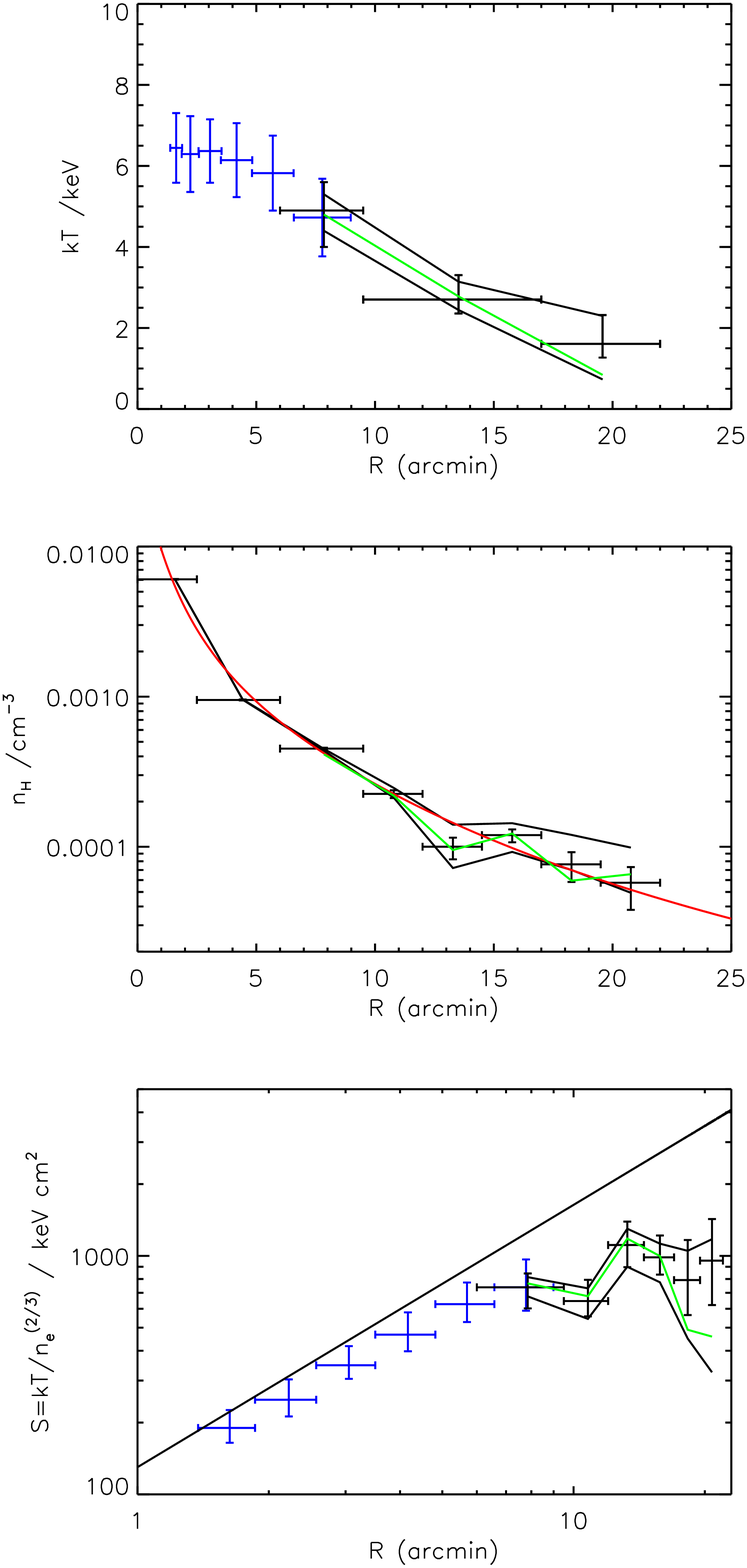,
width=\columnwidth }
       }
      \caption{Deprojected temperature, density and entropy profiles. Left
column is for the azimuthal average (excluding the north which has excess
emission and the south east which has a lower temperature and density). Right column
is for the south eastern direction which has a lower outskirts temperature
and flatter entropy profile. Solid black lines show one sigma systematic errors
from uncertainties in background modelling. The solid green line shows the
effect of
subtracting the stray light emission from the core. The red points overplotted
on the right are the \emph{XMM-Newton} temperatures from \citet{Bourdin2008} found for
the region away from the depression in temperature to the south east. The blue points
for the temperature and entropy on the left are the for the cold depression to
the south east found in \citet{Bourdin2008} and correlate well with the \emph{Suzaku} values
in the outskirts to the south east. For the density plots the solid red line is the best fit deprojected density profile found in \citet{Vikhlinin2006} using \emph{Chandra} data. For the entropy plots the solid black line is the r$^{1.1}$ relation from \citet{Voit2005} scaled to pass through the central three points of our Suzaku data on the left. The same profile is shown to the right to emphasise the lower entropy in the south east.}
      \label{deprojected_fits}
  \end{center}
\end{figure*}

Here we compute the systematic uncertainty of the background model on our
deprojected fits by producing 10000 random realisations of the background parameters
[the CXB normalisation, the 0.53 and 0.20 keV galactic \textsc{apec} component normalisations and
temperatures, and the NXB level (which has an uncertainty of $\pm$ 3.6 percent
from \citealt{Tawa2008}) distributed
by their expected variance and performing the deprojected fit for each. In this
way the systematic uncertainty in the background is folded into the
deprojection to give a complete propagation of the uncertainty in the
background to the uncertainty in the spectral fits. The one sigma
systematic errors on the temperature, density and entropy profiles resulting
from this process are shown for the deprojected fits in Fig.
\ref{deprojected_fits}. For the temperature the systematic error is roughly of
the same order as the statistical error (except in the outermost bin), but for the density it is the dominant
source of uncertainty in the outer 4 annuli.

The systematic errors are also calculated for the projected density and shown
in Fig. \ref{projected_densities}, showing that the excess to the north is still
statisically significant. Note that because these are projected fits the errors
are smaller than for the deprojected fits (because for the deprojected fits the error on the outer annuli is propagated to the error on the inner annuli). 

We then perform all the deprojected fits with the stray light level from the
central 6$'$ (as simulated by \textsc{xissim}) subtracted, and plot as the green line in
Fig. \ref{deprojected_fits} the best fit results with this stray light
subtracted. In all cases the effect of taking into account the stray light is
within the systematic error of the background modelling.

We vary the contamination level on the optical blocking filter (OBF) by $\pm$10 percent as described in \citet{Akamatsu2011}, using the \emph{Suzaku} ftool \textsc{xiscontamicalc} to modifty the ARFS, and find this to have a negligible affect on the fits.  

\section{Mass analysis}
\label{mass_analysis}

Assuming hydrostatic equilibrium and spherical symmetry the total mass interior to radius r is given by \citep{Vikhlinin2006}

\begin{equation}
\label{hydroeq}
M(<r)=-\mathrm{3.68}\times\mathrm{10}^{\mathrm{13}}M_{\odot}T(r)r\left(\frac{d \rm \:ln \:n_{\rm
    H}}{d \:\mathrm{ln \:r}} + \frac{d\rm\:ln\:T}{d\rm\:ln\:r}\right) \\
\end{equation}

where T is in keV and r is in Mpc. 
We find best fit temperature and density profiles by using Monte Carlo methods with 10000 trials, and use these profiles in equation \ref{hydroeq} to yield the gravitational mass profile 
and its 1 $\sigma$ error. The density profile is fit as a beta model of the form n$_{\mathrm{H}}$=n$_{\mathrm{0}}$(1+(r/r$_{\mathrm{cn}}$)$^{2}$)$^{-\beta}$, and the temperature profile is fit with the function T=T$_{\mathrm{0}}$(1+r/r$_{\mathrm{cT}}$)$^{\alpha}$ as used in \citet{Akamatsu2011} and \citet{Burns2010}. We do this for the azimuthally averaged profiles shown in the left hand column of Fig. \ref{deprojected_fits} (thus excluding the north due to its excess emission and the south east due to its lower temperature and entropy which means that it is likely out of hydrostatic equilibrium). From this we then calculate an estimate of r$_{200}$ (the radius within which the mean density is 200 times the critical density of the universe), which is widely used as an approximation to the virial radius. The
cumulative gas mass is calculated using the number density of the ICM derived from the \textsc{apec} normalisation and assuming spherical symmetry, allowing the gas mass fraction, f$_{gas}$, as a function of radius to be derived.

The slightly shallower and lower temperature profile we find compared to \citet{Vikhlinin2006}, together with the slightly higher density in the outskirts causes our total mass estimate to be lower, and gas mass estimate to be higher, making our measurement of the gas mass fraction in the outskirts higher than those presented in \citet{Vikhlinin2006}. This consequently means that our estimate of r$_{200}$ is smaller. We find M$_{500}$=7.2$^{+1.1}_{-1.2}$ $\times$ 10$^{14}$ M$_{\odot}$, which is consistent with the \citet{Vikhlinin2006} value of M$_{500}$=8.01$\pm{0.74}$ $\times$ 10$^{14}$ M$_{\odot}$. We find r$_{200}$=1.92$^{+0.11}_{-0.13}$ Mpc (corresponding to 22.0$^{+1.3}_{-1.4}$ arcmins), with M$_{200}$=8.0$^{+1.5}_{-1.5}$ $\times$ 10$^{14}$ M$_{\odot}$, which is smaller than the value obtained using the scaling relation of \citet{Arnaud2005} :

\begin{eqnarray}
\label{mass_scal_relation_Arnaud}
M_{200}/M_{\odot}&=&5.74\pm0.3 \times10^{14} \Big[\dfrac{kT}{5\mathrm{keV}}\Big]^{1.49 \pm 0.17}/E(z)^{1/2} \\
E(z)&=&[\Omega_{m}(1+z)^{3}+1-\Omega_{m}] \\
\end{eqnarray}

which yields M$_{200}$ =12.1$^{+0.7}_{-0.8}$ $\times$ 10 $^{14}$ M$_{\odot}$ and r$_{200}$=25.3$^{+0.6}_{-0.6}$ using the mean spectroscopic temperature of 8.47$\pm$0.09 keV from \emph{Chandra} obtained in \citet{Vikhlinin2006}. Part of this discrepancy is, as described in \citet{Bautz2009} (which similarly finds a lower mass and r$_{200}$ than derived using \emph{Chandra} results and scaling relations), likely due to the fact that \emph{Chandra} reports temperatures for hot clusters which 
are systematically higher than those from \emph{XMM-Newton}. This can be seen by comparing the \emph{XMM-Newton} temperature points in Fig. \ref{deprojected_fits} with the \emph{Chandra} points in Fig. \ref{temperature_compare}, and it is believed to be due to the uncertainty in the contamination on the \emph{Chandra} mirrors. Our \emph{Suzaku} results suggest a lower emission averaged spectroscopic temperature of 7.5$\pm$0.1 keV (which is consistent with the average temperature of 7.45$\pm$0.19 found with ASCA in \citealt{Fukazawa2004} and is more consistent with the \emph{XMM-Newton} temperatures), which when used in equation \ref{mass_scal_relation_Arnaud}, gives M$_{200}$=10.1$^{+0.6}_{-0.6}$  $\times$ 10$^{14}$ M$_{\odot}$ and r$_{200}$=23.8$^{+0.5}_{-0.5}$ arcmins, which are consistent with the values derived from our \emph{Suzaku} fits. 

\begin{figure}
  \begin{center}
    \leavevmode
      \epsfig{figure=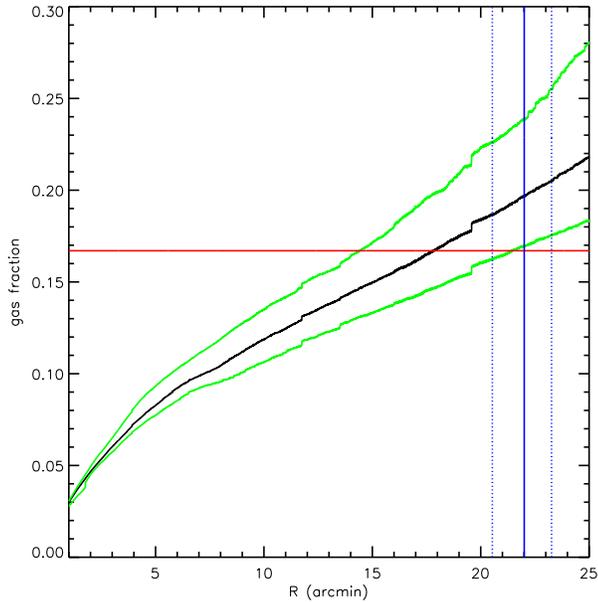,
        width=\linewidth}
             \caption{Gas mass fraction as a function of radius for Abell 2029. Black line shows the best fit and the green lines show the 1 $\sigma$ error ranges. The red horizontal line shows the cosmic mean baryon fraction as determined in \citet{Komatsu2011}. The solid blue vertical line shows our measurement of r$_{200}$ and the vertical blue dotted lines show the one sigma uncertainty in r$_{200}$. }
      \label{gasmassfraction}
  \end{center}
\end{figure}

The gas mass fraction is shown in Fig. \ref{gasmassfraction}, which rises with radius and reaches the cosmic mean baryon fraction (0.167, obtained from CMB data in \citealt{Komatsu2011}) near our calculated value of the virial radius. 

\section{Excess to the north.}
\label{Filament}

The projected density profile shows an excess to the North, suggesting the
existence of a filament connecting Abell 2029 to Abell 2033, or contaminating emission from Abell 2033 entering the field of view. The existence of
this structure has also been claimed in
\citet{Eckert2011b} who studied azimuthal variations in the surface brightness
profiles of 31 galaxy clusters using archival ROSAT PSPC data, and found a
significant excess connecting  Abell 2029 and
Abell 2033.

\begin{figure}
  \begin{center}

    \leavevmode
        \hbox{
       \epsfig{figure=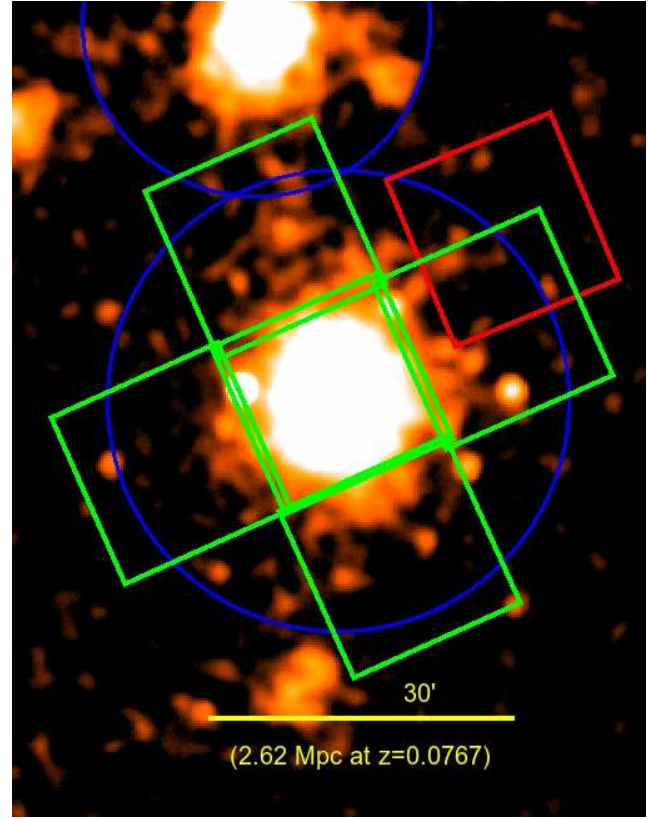,
width=\columnwidth }
       }
      \caption{Exposure corrected and background subtracted ROSAT PSPC image of
the Abell 2029/2033 system smoothed with a Gaussian kernel, showing indications
of excess emission to the north of Abell 2029 between the two clusters. The virial radii of the two clusters are shown by the blue circles.}
      \label{ROSAT_PSPCimage}
  \end{center}
\end{figure}

To check the presence of this excess we used a ROSAT PSPC pointing of the
Abell 2029 and Abell 2033 system (rp800249) and examined the surface brightness
profiles in the directions corresponding
to the arms of the cross (using the same regions and annuli binning as used for
our \emph{Suzaku} pointings). The ROSAT PSPC data were reduced using the ROSAT
Extended Source Analysis Software (ESAS, \citealt{Snowden1994}).
We follow the same method as \citet{Eckert2011a} to obtain a background
subtracted, exposure corrected counts image for the R47 band (corresponding to
0.44-2.04keV). The
resulting image is shown in Fig. \ref{ROSAT_PSPCimage}, and there is a clear
bright region connecting Abell 2029 to Abell 2033.

\begin{figure*}
  \begin{center}

    \leavevmode
        \hbox{
       \epsfig{figure=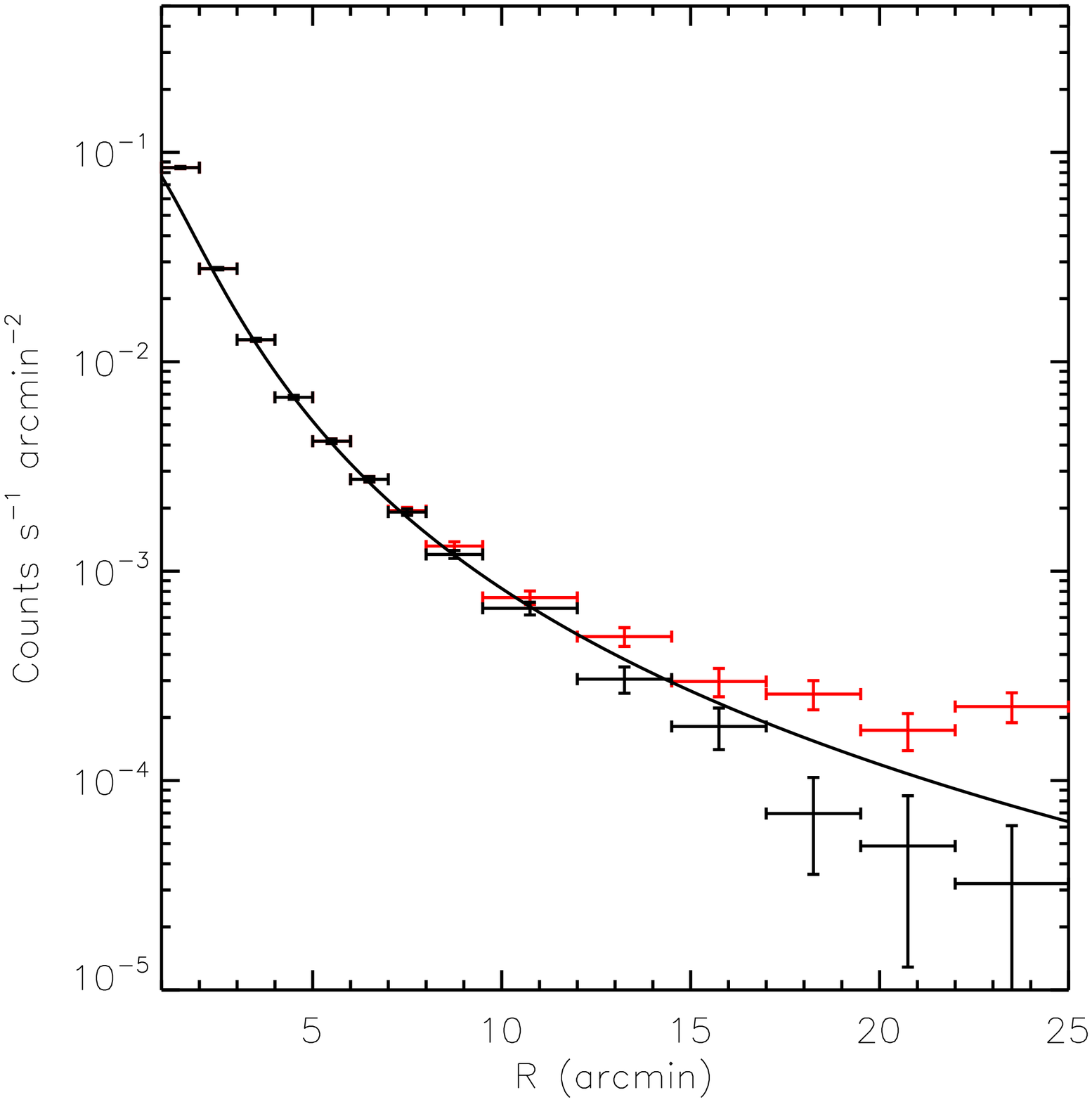,
width=\columnwidth }
              \epsfig{figure=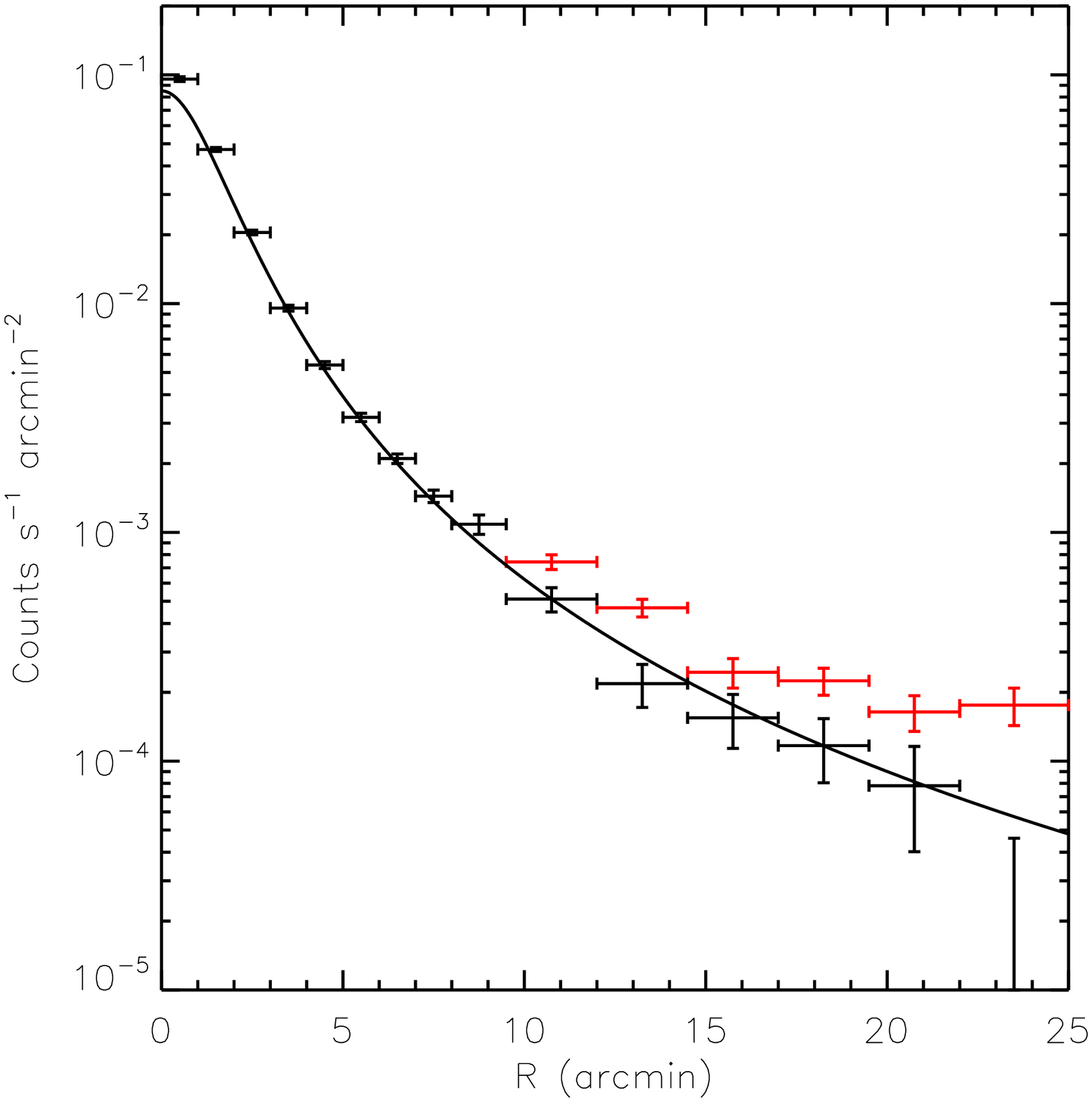, width=\columnwidth }

       }
      \caption{Left: Exposure corrected and background subtracted ROSAT PSPC
surface brightness profile for the same regions analysed with \emph{Suzaku} of the
Abell 2029/2033 system in the 0.4-2.0 keV band. Right:Vignetting corrected and
background subtracted \emph{Suzaku} surface brightness profile in the 0.7-2.0 keV band, averaged over the two front illuminated detectors (XIS0 and XIS3).
In both panels the black points are an azimuthal average over all directions
except the north and the red points are for the northern direction only, showing
a significant excess to the North with both satellites. The black solid line is the beta model found in S98 for the azimuthally averaged surface brightness profile between 164 arcsecs and 1000 arcsecs (16.7 arcmin). }
      \label{ROSATproj_sb}
  \end{center}
\end{figure*}

The resulting background subtracted projected ROSAT surface brightness profile is shown in Fig.
\ref{ROSATproj_sb}, where the error in the background has been added in quadrature to the error bars. We clearly see an excess to the north (red) deviating from
the beta law decrease of the azimuthal average of the other directions. This is
the only statistically significant deviation from azimuthal symmetry in the
surface brightness profiles. We also show the corresponding vignetting corrected and background subtracted projected surface
brightness profile obtained with \emph{Suzaku} in the 0.7-2.0keV band in Fig.
\ref{ROSATproj_sb} (where vignetting correction and background subtraction were performed following the method of \citealt{Bautz2009}), also showing the excess in the northern direction.  The
overplotted beta model is that obtained in S98 (which used different data
reduction methods) as the best fit to the azimuthally averaged profile between
164 arcsecs and 1000 arcsecs.

However another possibility is that the excess is caused due to overlapping
emission from
the outskirts of Abell 2029 and Abell 2033, with no need for excess emission
from a filament. Using the scaling relation, equation \ref{mass_scal_relation_Arnaud}, for
r$_{200}$  we find r$_{200}$ for Abell 2033 to be 1.5 Mpc
(17.0$'$) (using the mean temperature of 3.84 keV from ASCA found in \citealt{Fukazawa2004}). We found in section \ref{mass_analysis} that for Abell 2029 our \emph{Suzaku} results imply r$_{200}$=22.0$^{+1.3}_{-1.4}$ arcmins. These virial radii are overplotted as the
blue circles on Fig. \ref{ROSAT_PSPCimage}, and we see that the excess emission
correlates well with the region where we would expect to detect emission from the ICM of the outskirts of Abell 2033.

\subsection{Spectral analysis of northern excess}
\label{spectrum_excess}

The difference in redshift between Abell 2029 (z=0.0767) and Abell 2033
(z=0.081) corresponds to a distance of $\sim$ 20 Mpc (neglecting the effect of
possible peculiar velocities), meaning that if a filament exists between the
two then we are seeing it along
its length, increasing its surface brightness due to this preferential
geometry. Such a preferential geometry has already been exploited to study a
potential filament between Abell 222 and Abell 223
 \citep{Werner2008} which is viewed along its length, however as that system is
located at a much higher redshift (z=0.21) than the Abell 2029 and Abell 2033
system its surface brightness and area of extent
are both lower.

\begin{figure}
  \begin{center}

    \leavevmode
        \hbox{
       \epsfig{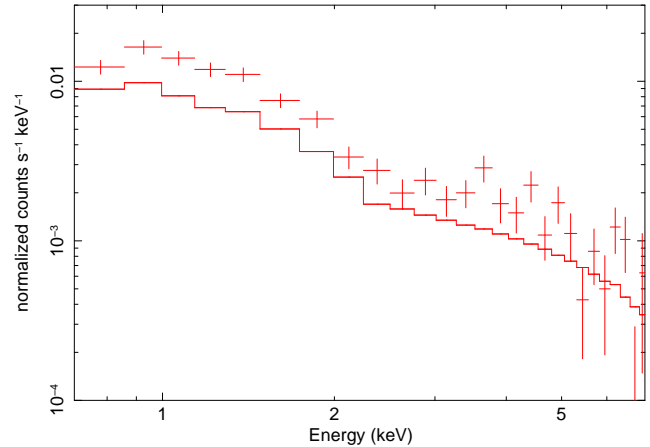}
       }
      \caption{Spectrum of excess emission to the north. Points show spectrum
of northern excess region. Line shows the azimuthal average spectrum fit for
all of the other directions.}
      \label{excess_spectrum}
  \end{center}
\end{figure}

To fit for the projected properties of the northern excess, we perform a fit to
the
19.5$'$-25.0$'$ region to the north, and simultaneously fit the same regions in
the other 4 directions which are tied to get
an azimuthal average, and use this as a background model. In this way we fit
the excess over the azimuthal average level to the north in the 19.5$'$-25.0$'$
region as an absorbed \textsc{apec} component. The excess of the spectrum over the
azimuthal average of the other directions is shown in Fig.
\ref{excess_spectrum}.
The normalisation of the excess \textsc{apec} component, taking into account systematic errors in the
background determination is  1.5$^{+0.5}_{-0.5}$ $\times$ 10 $^{-3}$ cm$^{-5}$, and the
temperature
is 3.6$^{+1.1}_{-1.2}$ keV). This temperature appears too high for the excess
to be caused by a WHIM filament connecting the two clusters (which would be
expected to have a temperature around 1 keV as this gas would not have been shock
heated due to accretion yet (\citealt{Werner2008}, \citealt{Dave2001}). The temperature is
more consistent with this emission originating from
the outskirts of Abell 2033. Assuming that the excess emission completely
originates from Abell 2033 and assuming spherical symmetry, the emission norm
of the excess corresponds to a density in the range 1.0-2.0 $\times$ 10 $^{-4}$
cm$^{-3}$ in the outskirts of Abell 2033, which seems reasonable.

Figures \ref{projected_densities} and \ref{ROSATproj_sb} show that the excess to the northern direction is present from a radius of 12$'$ outwards, but it only seems reasonable to attribute the excess at radii beyond 19.5$'$ as being due to overlapping emission from Abell 2033. The excess between 12$'$ and 19.5$'$ could have a number of causes. One is that the density to the north could genuinely be higher than that in other directions. Another is that we may be detecting emission from gas in a filamentary structure connecting Abell 2029 and Abell 2033.

%Assuming that all of the excess emission
%originates from the filament region shown in the Fig.
%\ref{ROSAT_PSPCimage}, and that the length is 20 Mpc, this emission measure
%corresponds to a density of 3.0$\times$10$^{-5}$ cm-3, which is
%consistent with the density range expected for such structures
%\citep{Dave2001}.

\subsection{Comparing to the spatial galaxy distribution}
\label{SDSS_data}

\begin{figure}
  \begin{center}
    \leavevmode
      \epsfig{figure=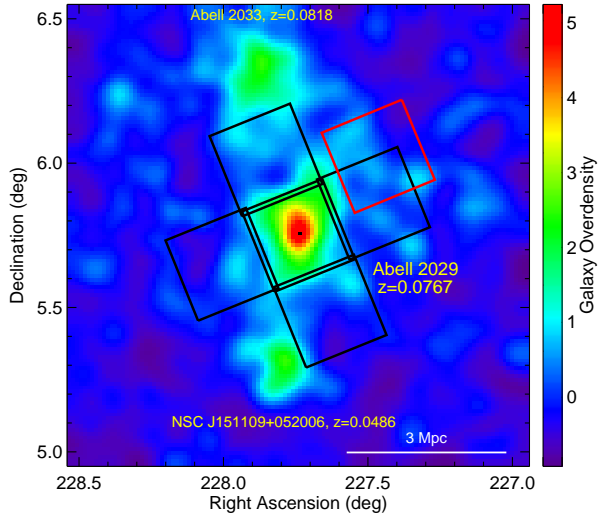,
        width=\linewidth}
             \caption{This shows the distribution of galaxies with magnitude
i$<$21.0 and photometric redshift $\vert z -z_{c} \vert <$0.016 using SDSS
data. We define the overdensity as $(\rho-\langle\rho\rangle) /
\langle\rho\rangle$ where $\langle\rho\rangle$ is the mean galaxy density
computed on the 16 deg$^{2}$ region centred on Abell 2029, and smooth the
overdensity plot using a Gaussian kernel with a FWHM of 500 kpc at the redshift
of Abell 2029. }
      \label{SDSS1}
  \end{center}
\end{figure}

To investigate the nature of the excess emission between Abell 2029 and Abell 2033 further, we constructed
the density field of galaxies around Abell 2029 using data from the
Sloan Digital Sky Survey (SDSS). Taking the photometric catalogue from
SDSS Data Release 7 \citep{Abazajian2009}, we select galaxies with magnitude
$i<21.0$ and a photometric
redshift within a range of $dz=0.032$ centred on the redshift of
Abell 2029, $z_{\rm c}=0.0767$. This range corresponds to the median
photometric redshift uncertainty for sources of this magnitude at
$z<0.1$. The general features of the density field are not highly
sensitive to the redshift range used, and the spectroscopic catalogue
produces similar results though with a lower sampling density. Though
the range exceeds the typical velocity dispersion of a cluster, we can
easily identify peaks in the density field associated with known
clusters nearby, as shown in Fig. \ref{SDSS1}. We plot the
overdensity of galaxies smoothed by a Gaussian with FWHM$=500$~kpc.

There is a clear excess in the galaxy density to the north which appears to originate from Abell 2033, consistent with our
hypothesis that we are seeing the overlap of the virial radii of Abell 2029 and
Abell 2033.

The foreground object NSC J151109+052006 is very faint in X-rays (it is visible as an extended source to the south in Fig. \ref{ROSAT_PSPCimage}) and lies out of the field of view of the south west pointing. We simulated the expected contribution from this source in the south west pointing using \textsc{xissim} and found it to be negligible.  

\section{Summary}
\label{summary}
We explore the ICM of the galaxy cluster Abell 2029 to higher radius than
before, with no statistically significant emission detected beyond 22$'$ (except for the northern pointing between Abell 2029 and Abell 2033), which
is 1.9Mpc, and is equal to our measurement of r$_{200}$. Excess emission is found to the north between Abell 2029
and Abell 2033 which is detected in both our \emph{Suzaku} observations and in ROSAT
PSPC observations, and the most consistent explanation for this excess emission
is that it originates from Abell 2033 and we are seeing the overlap of the
outskirts of the two clusters to the north.

We detect a cold feature to the SE extending out to the edge of the detected
cluster (22$'$) where the ICM is significantly colder than in the other
directions, consistent with the \emph{XMM-Newton} findings of \citet{Bourdin2008}
which found a temperature depression to the south east within the central 8$'$.
The substantial
point spread function does not allow this cold feature to be traced into the
central 2 annuli of the \emph{Suzaku} pointing. The lower temperature causes the entropy profile to be
lower and to flatten, indicating that the ICM in this direction is
significantly out of hydrostatic equilibrium. This may be the result of the
accretion of galaxy groups along the SE direction, which has disturbed the ICM.

Away from this cold feature, the azimuthally averaged temperatures and
densities are consistent with previous findings, with the entropy profile
rising in the outskirts but not as steeply as the r$^{1.1}$ powerlaw predicted
due to pure gravitational hierarchical structure formation, a result which has
also been found for other clusters [Abell 1795, \citep{Bautz2009}, the Perseus
cluster \citep{Simionescu2011}, and Abell 2142 \citep{Akamatsu2011}].

\section*{Acknowledgements}

SAW is supported by STFC, and ACF thanks the Royal Society. MRG acknowledges
support from an NSF Graduate Research Fellowship and NASA grant NNX10AR49G. This research has used data from the $Suzaku$
telescope, a joint mission between JAXA and NASA. We acknowledge the
use of data made available through the SDSS archive. Funding for the
Sloan Digital Sky Survey (SDSS) has been provided by the Alfred P.
Sloan Foundation, the Participating Institutions, the National Science
Foundation, the U.S. Department of Energy, the Japanese
Monbukagakusho, and the Max Planck Society. This research has made use
of the NASA/IPAC Extragalactic Database (NED) which is operated by the
Jet Propulsion Laboratory, California Institute of Technology, under
contract with the National Aeronautics and Space Administration.

\bibliographystyle{mn2e}
\bibliography{Abell2029_paper}

\begin{thebibliography}{}

\bibitem[\protect\citeauthoryear{{Abazajian}, {Adelman-McCarthy},
  {Ag{\"u}eros}, {Allam}, {Allende Prieto}, {An}, {Anderson}, {Anderson},
  {Annis}, {Bahcall} \& {et al.}}{{Abazajian} et~al.}{2009}]{Abazajian2009}
{Abazajian} K.~N.,  {Adelman-McCarthy} J.~K.,  {Ag{\"u}eros} M.~A.,  {Allam}
  S.~S.,  {Allende Prieto} C.,  {An} D.,  {Anderson} K.~S.~J.,  {Anderson}
  S.~F.,  {Annis} J.,  {Bahcall} N.~A.,    {et al.} 2009, \apjs, 182, 543

\bibitem[\protect\citeauthoryear{{Akamatsu}, {Hoshino}, {Ishisaki}, {Ohashi},
  {Sato}, {Takei} \& {Ota}}{{Akamatsu} et~al.}{2011}]{Akamatsu2011}
{Akamatsu} H.,  {Hoshino} A.,  {Ishisaki} Y.,  {Ohashi} T.,  {Sato} K.,
  {Takei} Y.,    {Ota} N.,  2011, \pasj, 63, 1019

\bibitem[\protect\citeauthoryear{{Allen}, {Rapetti}, {Schmidt}, {Ebeling},
  {Morris} \& {Fabian}}{{Allen} et~al.}{2008}]{Allen2008}
{Allen} S.~W.,  {Rapetti} D.~A.,  {Schmidt} R.~W.,  {Ebeling} H.,  {Morris}
  R.~G.,    {Fabian} A.~C.,  2008, \mnras, 383, 879

\bibitem[\protect\citeauthoryear{{Arnaud}, {Pointecouteau} \& {Pratt}}{{Arnaud}
  et~al.}{2005}]{Arnaud2005}
{Arnaud} M.,  {Pointecouteau} E.,    {Pratt} G.~W.,  2005, \aap, 441, 893

\bibitem[\protect\citeauthoryear{{Bautz}, {Miller}, {Sanders}, {Arnaud},
  {Mushotzky}, {Porter}, {Hayashida}, {Henry}, {Hughes}, {Kawaharada},
  {Makashima}, {Sato} \& {Tamura}}{{Bautz} et~al.}{2009}]{Bautz2009}
{Bautz} M.~W.,  {Miller} E.~D.,  {Sanders} J.~S.,  {Arnaud} K.~A.,  {Mushotzky}
  R.~F.,  {Porter} F.~S.,  {Hayashida} K.,  {Henry} J.~P.,  {Hughes} J.~P.,
  {Kawaharada} M.,  {Makashima} K.,  {Sato} M.,    {Tamura} T.,  2009, \pasj,
  61, 1117

\bibitem[\protect\citeauthoryear{{Bourdin} \& {Mazzotta}}{{Bourdin} \&
  {Mazzotta}}{2008}]{Bourdin2008}
{Bourdin} H.,  {Mazzotta} P.,  2008, \aap, 479, 307

\bibitem[\protect\citeauthoryear{{Burns}, {Skillman} \& {O'Shea}}{{Burns}
  et~al.}{2010}]{Burns2010}
{Burns} J.~O.,  {Skillman} S.~W.,    {O'Shea} B.~W.,  2010, \apj, 721, 1105

\bibitem[\protect\citeauthoryear{{Clarke}, {Blanton} \& {Sarazin}}{{Clarke}
  et~al.}{2004}]{Clarke2004}
{Clarke} T.~E.,  {Blanton} E.~L.,    {Sarazin} C.~L.,  2004, \apj, 616, 178

\bibitem[\protect\citeauthoryear{{Dav{\'e}}, {Cen}, {Ostriker}, {Bryan},
  {Hernquist}, {Katz}, {Weinberg}, {Norman} \& {O'Shea}}{{Dav{\'e}}
  et~al.}{2001}]{Dave2001}
{Dav{\'e}} R.,  {Cen} R.,  {Ostriker} J.~P.,  {Bryan} G.~L.,  {Hernquist} L.,
  {Katz} N.,  {Weinberg} D.~H.,  {Norman} M.~L.,    {O'Shea} B.,  2001, \apj,
  552, 473

\bibitem[\protect\citeauthoryear{{De Luca} \& {Molendi}}{{De Luca} \&
  {Molendi}}{2004}]{DeLuca2004}
{De Luca} A.,  {Molendi} S.,  2004, \aap, 419, 837

\bibitem[\protect\citeauthoryear{{Eckert}, {Molendi}, {Gastaldello} \&
  {Rossetti}}{{Eckert} et~al.}{2011}]{Eckert2011a}
{Eckert} D.,  {Molendi} S.,  {Gastaldello} F.,    {Rossetti} M.,  2011, \aap,
  529, A133+

\bibitem[\protect\citeauthoryear{{Eckert}, {Vazza}, {Ettori}, {Molendi},
  {Nagai}, {Lau}, {Roncarelli}, {Rossetti}, {Snowden} \&
  {Gastaldello}}{{Eckert} et~al.}{2011}]{Eckert2011b}
{Eckert} D.,  {Vazza} F.,  {Ettori} S.,  {Molendi} S.,  {Nagai} D.,  {Lau}
  E.~T.,  {Roncarelli} M.,  {Rossetti} M.,  {Snowden} S.~L.,    {Gastaldello}
  F.,  2011, ArXiv e-prints

\bibitem[\protect\citeauthoryear{{Einasto}, {Einasto}, {Tago}, {M{\"u}ller} \&
  {Andernach}}{{Einasto} et~al.}{2001}]{Einasto2001}
{Einasto} M.,  {Einasto} J.,  {Tago} E.,  {M{\"u}ller} V.,    {Andernach} H.,
  2001, \aj, 122, 2222

\bibitem[\protect\citeauthoryear{{Fukazawa}, {Makishima} \&
  {Ohashi}}{{Fukazawa} et~al.}{2004}]{Fukazawa2004}
{Fukazawa} Y.,  {Makishima} K.,    {Ohashi} T.,  2004, \pasj, 56, 965

\bibitem[\protect\citeauthoryear{{Gastaldello}, {Ettori}, {Balestra},
  {Brighenti}, {Buote}, {de Grandi}, {Ghizzardi}, {Gitti} \&
  {Tozzi}}{{Gastaldello} et~al.}{2010}]{Gastaldello2010}
{Gastaldello} F.,  {Ettori} S.,  {Balestra} I.,  {Brighenti} F.,  {Buote}
  D.~A.,  {de Grandi} S.,  {Ghizzardi} S.,  {Gitti} M.,    {Tozzi} P.,  2010,
  \aap, 522, A34+

\bibitem[\protect\citeauthoryear{{George}, {Fabian}, {Sanders}, {Young} \&
  {Russell}}{{George} et~al.}{2009}]{George2009}
{George} M.~R.,  {Fabian} A.~C.,  {Sanders} J.~S.,  {Young} A.~J.,    {Russell}
  H.~R.,  2009, \mnras, 395, 657

\bibitem[\protect\citeauthoryear{{Hoshino}, {Patrick Henry}, {Sato},
  {Akamatsu}, {Yokota}, {Sasaki}, {Ishisaki}, {Ohashi}, {Bautz}, {Fukazawa},
  {Kawano}, {Furuzawa}, {Hayashida}, {Tawa}, {Hughes}, {Kokubun} \&
  {Tamura}}{{Hoshino} et~al.}{2010}]{Hoshino2010}
{Hoshino} A.,  {Patrick Henry} J.,  {Sato} K.,  {Akamatsu} H.,  {Yokota} W.,
  {Sasaki} S.,  {Ishisaki} Y.,  {Ohashi} T.,  {Bautz} M.,  {Fukazawa} Y.,
  {Kawano} N.,  {Furuzawa} A.,  {Hayashida} K.,  {Tawa} N.,  {Hughes} J.~P.,
  {Kokubun} M.,    {Tamura} T.,  2010, \pasj, 62, 371

\bibitem[\protect\citeauthoryear{{Humphrey}, {Buote}, {Brighenti}, {Flohic},
  {Gastaldello} \& {Mathews}}{{Humphrey} et~al.}{2011}]{Humphrey2011}
{Humphrey} P.~J.,  {Buote} D.~A.,  {Brighenti} F.,  {Flohic} H.~M.~L.~G.,
  {Gastaldello} F.,    {Mathews} W.~G.,  2011, ArXiv e-prints

\bibitem[\protect\citeauthoryear{{Ishisaki}, {Maeda}, {Fujimoto}, {Ozaki},
  {Ebisawa}, {Takahashi}, {Ueda}, {Ogasaka}, {Ptak}, {Mukai} \&
  {Hamaguchi}}{{Ishisaki} et~al.}{2007}]{Ishisaki2007}
{Ishisaki} Y.,  {Maeda} Y.,  {Fujimoto} R.,  {Ozaki} M.,  {Ebisawa} K.,
  {Takahashi} T.,  {Ueda} Y.,  {Ogasaka} Y.,  {Ptak} A.,  {Mukai} K.,
  {Hamaguchi} K.,  2007, \pasj, 59, 113

\bibitem[\protect\citeauthoryear{{Kalberla}, {Burton}, {Hartmann}, {Arnal},
  {Bajaja}, {Morras} \& {P{\"o}ppel}}{{Kalberla} et~al.}{2005}]{LAB2005}
{Kalberla} P.~M.~W.,  {Burton} W.~B.,  {Hartmann} D.,  {Arnal} E.~M.,  {Bajaja}
  E.,  {Morras} R.,    {P{\"o}ppel} W.~G.~L.,  2005, \aap, 440, 775

\bibitem[\protect\citeauthoryear{{Kawaharada}, {Okabe}, {Umetsu}, {Takizawa},
  {Matsushita}, {Fukazawa}, {Hamana}, {Miyazaki}, {Nakazawa} \&
  {Ohashi}}{{Kawaharada} et~al.}{2010}]{Kawaharada2010}
{Kawaharada} M.,  {Okabe} N.,  {Umetsu} K.,  {Takizawa} M.,  {Matsushita} K.,
  {Fukazawa} Y.,  {Hamana} T.,  {Miyazaki} S.,  {Nakazawa} K.,    {Ohashi} T.,
  2010, \apj, 714, 423

\bibitem[\protect\citeauthoryear{{Komatsu}, {Smith}, {Dunkley}, {Bennett},
  {Gold}, {Hinshaw}, {Jarosik}, {Larson} \& {Nolta}}{{Komatsu}
  et~al.}{2011}]{Komatsu2011}
{Komatsu} E.,  {Smith} K.~M.,  {Dunkley} J.,  {Bennett} C.~L.,  {Gold} B.,
  {Hinshaw} G.,  {Jarosik} N.,  {Larson} D.,    {Nolta} M.~R.,  2011, \apjs,
  192, 18

\bibitem[\protect\citeauthoryear{{Kushino}, {Ishisaki}, {Morita}, {Yamasaki},
  {Ishida}, {Ohashi} \& {Ueda}}{{Kushino} et~al.}{2002}]{Kushino2002}
{Kushino} A.,  {Ishisaki} Y.,  {Morita} U.,  {Yamasaki} N.~Y.,  {Ishida} M.,
  {Ohashi} T.,    {Ueda} Y.,  2002, \pasj, 54, 327

\bibitem[\protect\citeauthoryear{{Lewis}, {Stocke} \& {Buote}}{{Lewis}
  et~al.}{2002}]{Lewis2002}
{Lewis} A.~D.,  {Stocke} J.~T.,    {Buote} D.~A.,  2002, \apjl, 573, L13

\bibitem[\protect\citeauthoryear{{Miyoshi}, {Tanaka}, {Yoshimura}, {Yamashita},
  {Furuzawa}, {Futamura} \& {Hudaverdi}}{{Miyoshi} et~al.}{2005}]{Miyoshi2005}
{Miyoshi} S.,  {Tanaka} N.,  {Yoshimura} M.,  {Yamashita} K.,  {Furuzawa} A.,
  {Futamura} T.,    {Hudaverdi} M.,  2005, Advances in Space Research, 36, 752

\bibitem[\protect\citeauthoryear{{Molendi} \& {De Grandi}}{{Molendi} \& {De
  Grandi}}{1999}]{Molendi1999}
{Molendi} S.,  {De Grandi} S.,  1999, \aap, 351, L41

\bibitem[\protect\citeauthoryear{{Moretti}, {Campana}, {Lazzati} \&
  {Tagliaferri}}{{Moretti} et~al.}{2003}]{Moretti2003}
{Moretti} A.,  {Campana} S.,  {Lazzati} D.,    {Tagliaferri} G.,  2003, \apj,
  588, 696

\bibitem[\protect\citeauthoryear{{Moretti}, {Gastaldello}, {Ettori} \&
  {Molendi}}{{Moretti} et~al.}{2011}]{Moretti2011}
{Moretti} A.,  {Gastaldello} F.,  {Ettori} S.,    {Molendi} S.,  2011, \aap,
  528, A102+

\bibitem[\protect\citeauthoryear{{Moretti}, {Pagani}, {Cusumano}, {Campana},
  {Perri}, {Abbey}, {Ajello}, {Beardmore}, {Burrows}, {Chincarini}, {Godet},
  {Guidorzi}, {Hill}, {Kennea}, {Nousek}, {Osborne} \& {Tagliaferri}}{{Moretti}
  et~al.}{2009}]{Moretti2009}
{Moretti} A.,  {Pagani} C.,  {Cusumano} G.,  {Campana} S.,  {Perri} M.,
  {Abbey} A.,  {Ajello} M.,  {Beardmore} A.~P.,  {Burrows} D.,  {Chincarini}
  G.,  {Godet} O.,  {Guidorzi} C.,  {Hill} J.~E.,  {Kennea} J.,  {Nousek} J.,
  {Osborne} J.~P.,    {Tagliaferri} G.,  2009, \aap, 493, 501

\bibitem[\protect\citeauthoryear{{Reiprich}, {Hudson}, {Zhang}, {Sato},
  {Ishisaki}, {Hoshino}, {Ohashi}, {Ota} \& {Fujita}}{{Reiprich}
  et~al.}{2009}]{Reiprich2009}
{Reiprich} T.~H.,  {Hudson} D.~S.,  {Zhang} Y.,  {Sato} K.,  {Ishisaki} Y.,
  {Hoshino} A.,  {Ohashi} T.,  {Ota} N.,    {Fujita} Y.,  2009, \aap, 501, 899

\bibitem[\protect\citeauthoryear{{Sarazin}, {Wise} \& {Markevitch}}{{Sarazin}
  et~al.}{1998}]{Sarazin1998}
{Sarazin} C.~L.,  {Wise} M.~W.,    {Markevitch} M.~L.,  1998, \apj, 498, 606

\bibitem[\protect\citeauthoryear{{Serlemitsos}, {Soong}, {Chan}, {Okajima},
  {Lehan}, {Maeda} \& {Itoh}}{{Serlemitsos} et~al.}{2007}]{Serlemitsos2007}
{Serlemitsos} P.~J.,  {Soong} Y.,  {Chan} K.-W.,  {Okajima} T.,  {Lehan} J.~P.,
   {Maeda} Y.,    {Itoh} K.,  2007, \pasj, 59, 9

\bibitem[\protect\citeauthoryear{{Simionescu}, {Allen}, {Mantz} \&
  {Werner}}{{Simionescu} et~al.}{2011}]{Simionescu2011conf}
{Simionescu} A.,  {Allen} S.,  {Mantz} A.,    {Werner} N.,  2011, in
  {J.-U.~Ness \& M.~Ehle} ed., The X-ray Universe 2011 {Baryons at the edge of
  the X-ray brightest galaxy cluster}.
p.~152

\bibitem[\protect\citeauthoryear{{Simionescu}, {Allen}, {Mantz}, {Werner},
  {Takei}, {Morris}, {Fabian}, {Sanders}, {Nulsen}, {George} \&
  {Taylor}}{{Simionescu} et~al.}{2011}]{Simionescu2011}
{Simionescu} A.,  {Allen} S.~W.,  {Mantz} A.,  {Werner} N.,  {Takei} Y.,
  {Morris} R.~G.,  {Fabian} A.~C.,  {Sanders} J.~S.,  {Nulsen} P.~E.~J.,
  {George} M.~R.,    {Taylor} G.~B.,  2011, Science, 331, 1576

\bibitem[\protect\citeauthoryear{{Snowden}, {Collier} \& {Kuntz}}{{Snowden}
  et~al.}{2004}]{Snowden2004}
{Snowden} S.~L.,  {Collier} M.~R.,    {Kuntz} K.~D.,  2004, \apj, 610, 1182

\bibitem[\protect\citeauthoryear{{Snowden}, {McCammon}, {Burrows} \&
  {Mendenhall}}{{Snowden} et~al.}{1994}]{Snowden1994}
{Snowden} S.~L.,  {McCammon} D.,  {Burrows} D.~N.,    {Mendenhall} J.~A.,
  1994, \apj, 424, 714

\bibitem[\protect\citeauthoryear{{Snowden}, {Mushotzky}, {Kuntz} \&
  {Davis}}{{Snowden} et~al.}{2008}]{Snowden2008}
{Snowden} S.~L.,  {Mushotzky} R.~F.,  {Kuntz} K.~D.,    {Davis} D.~S.,  2008,
  \aap, 478, 615

\bibitem[\protect\citeauthoryear{{Tawa}, {Hayashida}, {Nagai}, {Nakamoto},
  {Tsunemi}, {Yamaguchi}, {Ishisaki}, {Miller}, {Mizuno}, {Dotani}, {Ozaki} \&
  {Katayama}}{{Tawa} et~al.}{2008}]{Tawa2008}
{Tawa} N.,  {Hayashida} K.,  {Nagai} M.,  {Nakamoto} H.,  {Tsunemi} H.,
  {Yamaguchi} H.,  {Ishisaki} Y.,  {Miller} E.~D.,  {Mizuno} T.,  {Dotani} T.,
  {Ozaki} M.,    {Katayama} H.,  2008, \pasj, 60, 11

\bibitem[\protect\citeauthoryear{{Uson}, {Boughn} \& {Kuhn}}{{Uson}
  et~al.}{1991}]{Uson1991}
{Uson} J.~M.,  {Boughn} S.~P.,    {Kuhn} J.~R.,  1991, \apj, 369, 46

\bibitem[\protect\citeauthoryear{{Vikhlinin}, {Kravtsov}, {Forman}, {Jones},
  {Markevitch}, {Murray} \& {Van Speybroeck}}{{Vikhlinin}
  et~al.}{2006}]{Vikhlinin2006}
{Vikhlinin} A.,  {Kravtsov} A.,  {Forman} W.,  {Jones} C.,  {Markevitch} M.,
  {Murray} S.~S.,    {Van Speybroeck} L.,  2006, \apj, 640, 691

\bibitem[\protect\citeauthoryear{{Vikhlinin}, {Kravtsov}, {Burenin}, {Ebeling},
  {Forman}, {Hornstrup}, {Jones}, {Murray}, {Nagai}, {Quintana} \&
  {Voevodkin}}{{Vikhlinin} et~al.}{2009}]{Vikhlinin2009}
{Vikhlinin} A.,  {Kravtsov} A.~V.,  {Burenin} R.~A.,  {Ebeling} H.,  {Forman}
  W.~R.,  {Hornstrup} A.,  {Jones} C.,  {Murray} S.~S.,  {Nagai} D.,
  {Quintana} H.,    {Voevodkin} A.,  2009, \apj, 692, 1060

\bibitem[\protect\citeauthoryear{{Vikhlinin}, {Markevitch}, {Murray}, {Jones},
  {Forman} \& {Van Speybroeck}}{{Vikhlinin} et~al.}{2005}]{Vikhlinin2005}
{Vikhlinin} A.,  {Markevitch} M.,  {Murray} S.~S.,  {Jones} C.,  {Forman} W.,
   {Van Speybroeck} L.,  2005, \apj, 628, 655

\bibitem[\protect\citeauthoryear{{Voit}, {Kay} \& {Bryan}}{{Voit}
  et~al.}{2005}]{Voit2005}
{Voit} G.~M.,  {Kay} S.~T.,    {Bryan} G.~L.,  2005, \mnras, 364, 909

\bibitem[\protect\citeauthoryear{{Werner}, {Finoguenov}, {Kaastra},
  {Simionescu}, {Dietrich}, {Vink} \& {B{\"o}hringer}}{{Werner}
  et~al.}{2008}]{Werner2008}
{Werner} N.,  {Finoguenov} A.,  {Kaastra} J.~S.,  {Simionescu} A.,  {Dietrich}
  J.~P.,  {Vink} J.,    {B{\"o}hringer} H.,  2008, \aap, 482, L29

\end{thebibliography}

%\clearpage

\appendix
\section[]{Fitting for the galactic X-ray background}
\label{sec:appendix_gal}

\begin{figure*}
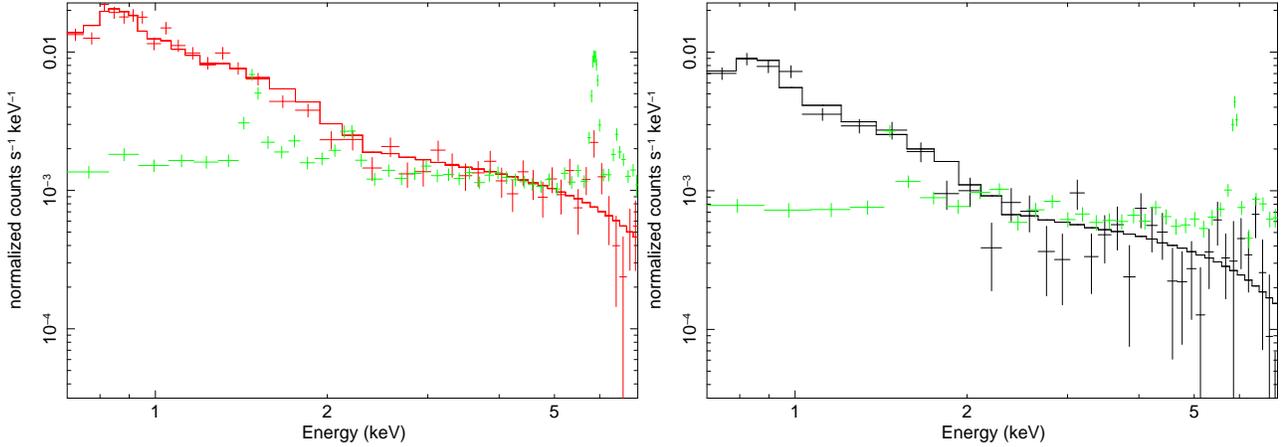

  \begin{center}

    \leavevmode
        \hbox{
       \epsfig{figure=FigA1a.eps, height=\columnwidth ,
angle=-90}
       \epsfig{figure=FigA1b.eps, height=\columnwidth ,
angle=-90}
       }
      \caption{Fitting to the \emph{Suzaku} background regions in the NW offset
pointing. Left shows the fit to the 25$'$-29$'$ annulus, and right shows the
fit to the 29$'$ to 34$'$ annulus using. Overplotted in green in the NXB level. Note that the areas of extraction for the two spectra are different, thus explaining the different normalised count rates on the y axis. 
}
      \label{Suz_bkg}
  \end{center}
\end{figure*}

\begin{figure*}
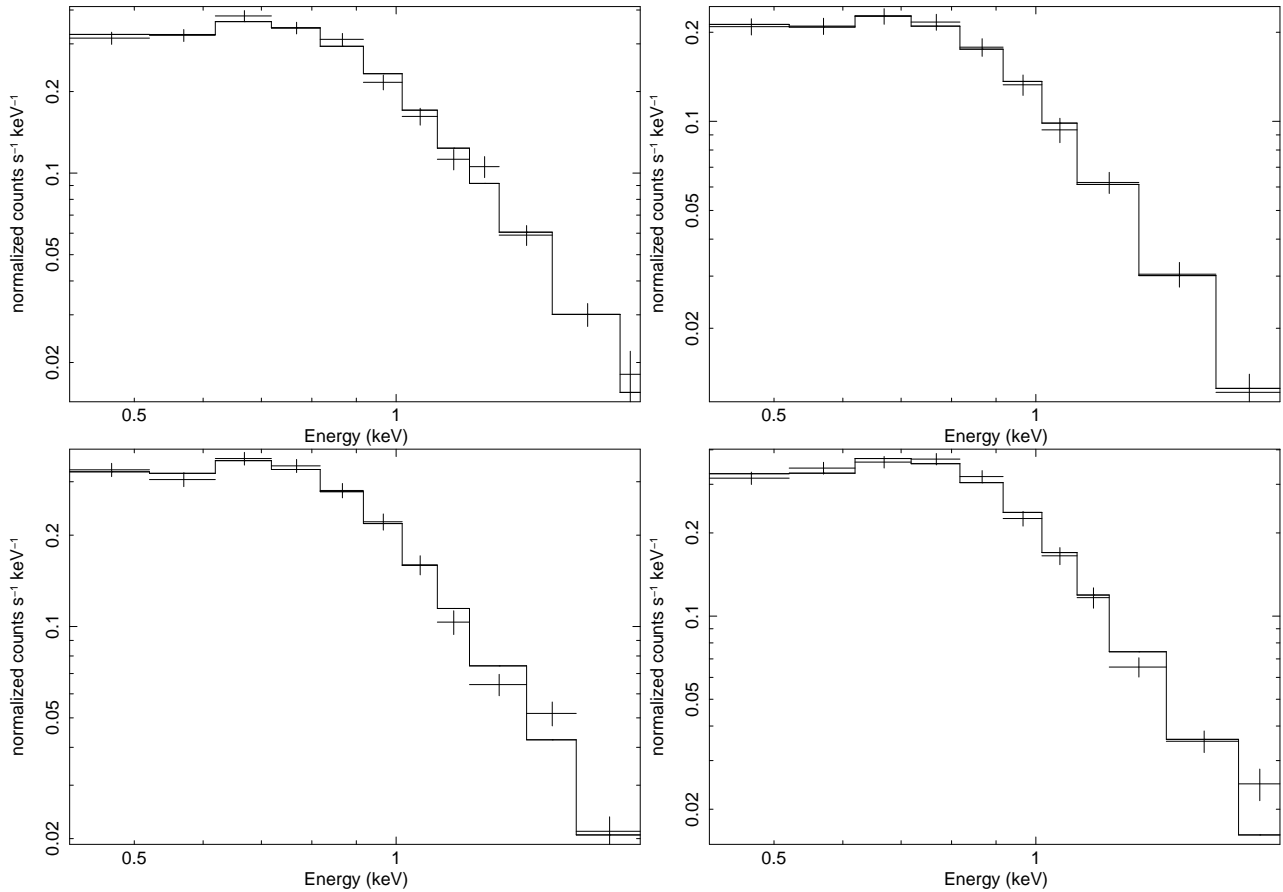

  \begin{center}

%    \leavevmode
         \hbox{
       \epsfig{figure=FigA2a.eps, height=\columnwidth , angle=-90}
       \epsfig{figure=FigA2b.eps, height=\columnwidth , angle=-90}
       }
         \hbox{
       \epsfig{figure=FigA2c.eps, height=\columnwidth , angle=-90}
       \epsfig{figure=FigA2d.eps, height=\columnwidth , angle=-90}
       }
      \caption{Fitting to ROSAT PSPC background regions in an annulus between
30$'$ and 50$'$ using the same model found using the \emph{Suzaku} background
pointings. Top left is for the NW, top right is for the NE, bottom left is for
the SE and bottom right is for the SW. These fits give the ranges in the
normalisations of the background parameters shown in table \ref{GALvariations}
which are used to understand the expected spatial variation of the soft X-ray
background. Note that the areas of extraction for the four spectra are different, thus explaining the different normalised count rates on the y axis. }
      \label{ROSAT_bkg}
  \end{center}
\end{figure*}

\clearpage

\section[]{Spectral Fitting}
\label{sec:appendix}

%Here we present a representative sample of the spectra used (which have been
%binned in energy with a minimum of 20 counts in each bin) and the best fitting
%models for each annulus and the background region. The spectra are well fitted
%by an absorbed \textsc{apec} model respresenting the cluster emission, added to
%the background model described in section \ref{backgroundmodelling}.

\begin{figure*}
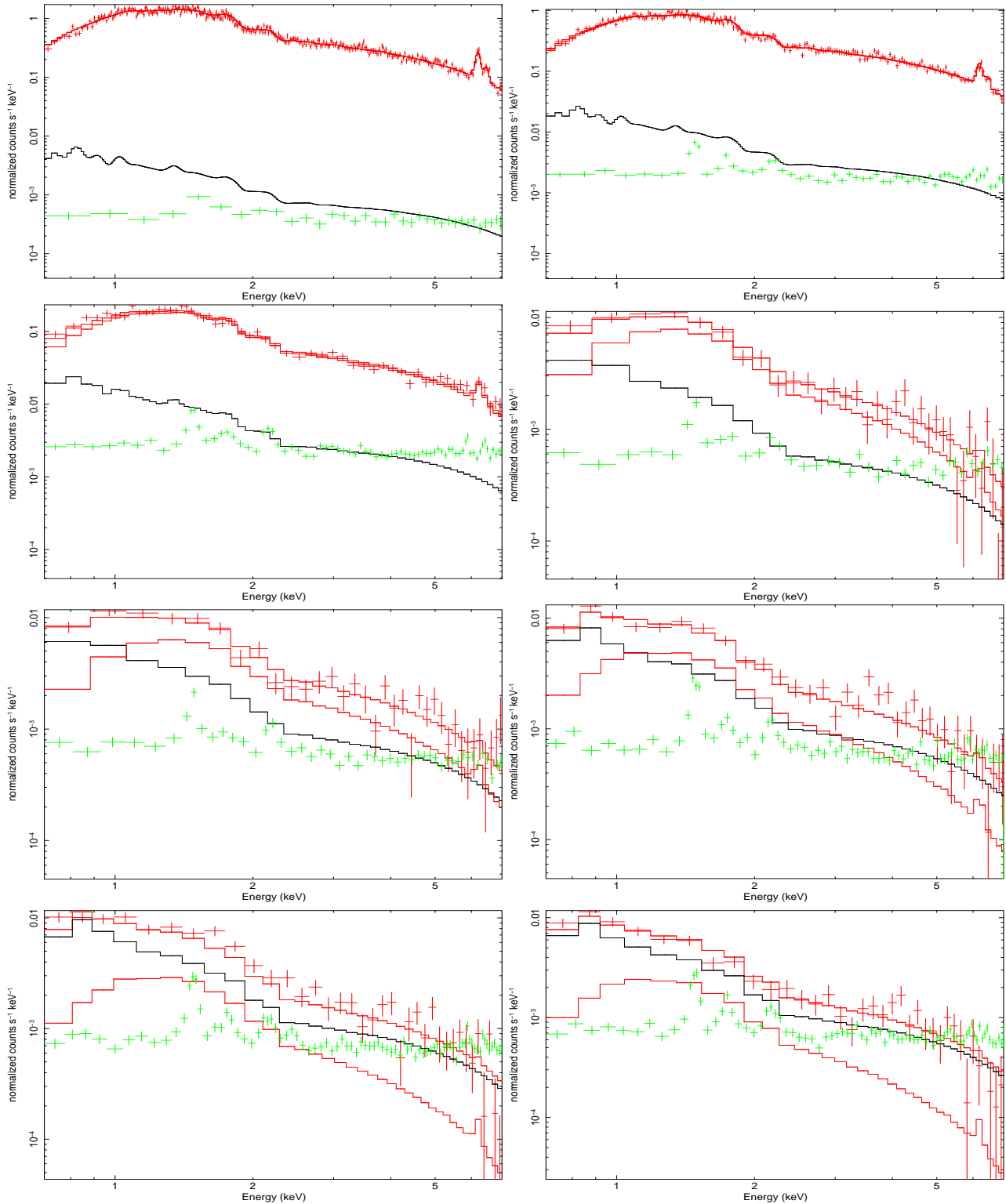

  \begin{center}
         \hbox{
       \epsfig{figure=FigB1a.ps,
height=\columnwidth ,width=0.6\columnwidth, angle=-90}
       \epsfig{figure=FigB1b.ps,
height=\columnwidth ,width=0.6\columnwidth, angle=-90}
       }
         \hbox{
       \epsfig{figure=FigB1c.ps,
height=\columnwidth ,width=0.6\columnwidth, angle=-90}
       \epsfig{figure=FigB1d.ps,
height=\columnwidth ,width=0.6\columnwidth, angle=-90}
       }
                \hbox{
       \epsfig{figure=FigB1e.ps,
height=\columnwidth ,width=0.6\columnwidth, angle=-90}
       \epsfig{figure=FigB1f.ps,
height=\columnwidth ,width=0.6\columnwidth, angle=-90}
       }
         \hbox{
       \epsfig{figure=FigB1g.ps,
height=\columnwidth ,width=0.6\columnwidth, angle=-90}
       \epsfig{figure=FigB1h.ps,
height=\columnwidth ,width=0.6\columnwidth, angle=-90}
       }
      \caption{Spectral fitting to the XIS0 and XIS3 data simultaneously. The spectra are for the azimuthally averaged fit (excluding the north and the south east) used to obtain the results in the left column of Fig. \ref{deprojected_fits}, but here we show the projected fits for simplicity. In reading order the spectra correspond to increasing radii annuli, which are between   0.0$'$-2.5$'$, 2.5$'$-6.0$'$, 6.0$'$-9.5$'$, 9.5$'$-12.0$'$,
12.0$'$-14.5$'$, 14.5$'$-17.0$'$, 17.0$'$-19.5$'$ and 19.5$'$-22.0$'$. The
central 3 annuli are integrated across all azimuth and are from the central pointing only. The red lines through the points represent the best fits (background plus
cluster emission), while the lower red line shows the excess over the background level (shown as the black solid line) modelled in these projected fits as an absorbed \textsc{apec} component. The NXB level is overplotted in green. }
      \label{spectra}
  \end{center}
\end{figure*}

\end{document}